\documentclass[a4paper,twoside]{article}
\newcommand{\affil}[1]{$^{\rm #1}$}
\usepackage[authoryear]{natbib}
\bibpunct{(}{)}{;}{a}{}{,}
\usepackage{graphicx}
\date{} 
\newcommand{\kms}{\mbox{km\,s$^{-1}$}}
\newcommand\HII{H~{\sc ii}}
\title{\large\bf\flushleft Precise Positions of Methanol Masers}
\author{\parbox{\textwidth}{\flushleft 
\vspace{-0.5cm}
{\it J.L.Caswell\affil{A,B}}\\
\vspace{0.4cm}
{\small \affil{A}\,ATNF, CSIRO, PO Box 76, Epping, NSW, Australia, 2121}\\
{\small \affil{B}\,Email: James.Caswell@csiro.au}}}
\begin{document}
\twocolumn[
\begin{changemargin}{.8cm}{.5cm}
\begin{minipage}{.9\textwidth}
\vspace{-1cm}
\maketitle
%
%
\small{\bf Abstract:}
The Australia Telescope Compact Array (ATCA) has been used to determine 
positions for many southern methanol maser sites, with accuracy better 
than 1 arcsec.  The results are presented here as a catalogue of more than 
350 distinct sites, some of them new discoveries, and many others with 
positional precision 10 times better than 
existing published values.  Clusters of 2 or 3 sites are occasionally 
found to account for single previously listed sources.  This in turn 
reveals that the velocity range for each individual site is sometimes 
smaller than that of the originally tabulated (blended) source.  Only a 
handful of examples then remain with a velocity range of more than 16 
\kms\  at a single compact (less than 2 arcsec)  site.  The precise 
methanol positions now allow apparent coincidences with OH masers to be 
confidently accepted or rejected;  this has led 
to the important conclusion that, where a 1665-MHz OH maser lies in a 
massive star formation region, at more than 80 percent of the OH sites 
there is a precisely coincident methanol maser. 
The methanol precision achieved here will also allow clear comparisons 
with likely associated IR sources when the next generation of far-IR 
surveys produce precise positions.


\medskip
\medskip
\end{minipage}
\end{changemargin}
]
\small

\section{Introduction}

Over the past decade, methanol maser emission at the 6668-MHz transition 
has become recognised as a valuable tracer of young stellar objects - the 
sites where massive stars have recently formed but are not directly 
detectable owing to their obscuring mantle of dust and molecules.  

Existing work has discovered a large number of methanol masers in our 
Galaxy by an inhomogeneous mixture of targeted searches (especially 
towards OH masers, and IR sources, e.g. Caswell et al. 1995a), and 
unbiased surveys towards some portions of the Galactic plane (e.g. 
Ellingsen et al. 1996).  However, there is a need to consolidate 
this work and provide accurate positions for the known sources in 
preparation for a new sensitive search for methanol masers that is 
currently being conducted with the Parkes radio telescope (Green et al. 
2009a).  

\section{Observations and data reduction}

The methanol maser observations described here were obtained with the ATCA 
in many sessions since 1993 February, chiefly in any of the 
four standard `6-km' configurations (instantaneously yielding 15 baselines 
ranging from 76 to 6000 m).  The correlator was configured to give a 
2048-channel spectrum across a 4-MHz 
bandwidth for each of the 2 orthogonal linear polarizations.  Typically, a 
target was observed for at least four periods of several minutes each, 
within a 10-hour timespan.  Targets selected for study included methanol 
sites with 
positions known only approximately from single dish observations, and the 
positions of some new OH masers (chiefly from Caswell 1998) not previously 
searched for methanol.  Some new serendiptious discoveries, made while 
investigating the chosen targets, are also reported. The masers selected 
for observation lie primarily in the Galactic longitude range 232$^\circ$ 
through 360$^\circ$ to 16$^\circ$ which is the region covered by extensive 
southern observations of OH masers (Caswell 1998).  Indeed, a major 
objective was to ensure that the precise position was obtained for all 
methanol masers that appeared to have a 
nearby maser counterpart of OH at either (or both) 1665 (or 1667) MHz and 
6035 MHz.  The region investigated was expanded to include    
some additional targets that lie between longitudes 
16$^\circ$ and 50$^\circ$, and between 188$^\circ$ and 232$^\circ$; these 
extensions were prompted by the absence until 
recently of a northern hemisphere instrument able to efficiently perform 
such measurements.  The procedures for observing and 
data reduction closely follow those of Caswell (1996a, 1996b, 1997).

\section{Results}

The synthesised beamsize of approximately 2 arcsec enables not only an 
accurate position measurement for the brightest maser feature in the 
spectrum of the target, but also enables mapping of the maser spot 
distribution.  Past studies have shown that the maser spots are 
generally confined to groups with typical maximum extent of less than 1 
arcsec (Caswell 1997; Forster \& Caswell 1989).  Where 
distances are known, the calculated linear extent rarely exceeds 30 mpc 
(=6000 au).  Sometimes the maps of maser spot positions show a cluster of 
two or more maser groups, with separations clearly exceeding group sizes.  
It is likely that each compact maser group within such a cluster has an 
embedded embryonic massive star as its source of excitation.




\begin{table*}[h]
\begin{center}
\caption{Methanol maser positions: \textit{- p1 of 8}}
\label{tableexample}
\begin{tabular}{lllrrccl}
\hline Galactic name & RA(2000) & Dec(2000) & I(pk) & v(pk) & 
v(range) & OH? & refs, epoch $^a$ \\
\ (~~l,~~~~~~~b) & (h~~m~~~s) & ~~$^\circ$~~~ '~~~~" & (Jy) & (\kms) 
& (\kms) & & \\
\hline
188.946+0.886	&	06 08 53.32	&	 +21 38 29.1	&	495	&	 +11	&	 \(-\)4, +12	&		&	99feb	\\
189.030+0.783	&	06 08 40.65	&	 +21 31 07.0	&	17	&	 +9	&	 +8, +10	&		&	99feb	\\
189.778+0.345	&	06 08 35.28	&	 +20 39 06.7	&	15	&	 +6	&	 +2, +6	&		&	99feb	\\
192.600\(-\)0.048	&	06 12 53.99	&	 +17 59 23.7	&	72	&	 +5	&	 +2, +6	&		&	99feb	\\
196.454\(-\)1.677	&	06 14 37.03	&	 +13 49 36.6	&	61	&	 +15	&	 +13, +16	&		&	99feb	\\
213.705\(-\)12.597	&	06 07 47.85	&	 \(-\)06 22 55.2	&	337	&	 +12	&	 +8, +13	&		&	99feb	\\
232.621+0.996	&	07 32 09.79	&	 \(-\)16 58 12.4	&	162	&	 +22.7	&	 +21, +24	&	c98	&	96dec	\\
263.250+0.514	&	08 48 47.84	&	 \(-\)42 54 28.4	&	60	&	 +12.3	&	 +11, +15	&	c98	&	96dec	\\
264.289+1.469	&	08 56 26.80	&	 \(-\)43 05 42.1	&	0.4	&	 +9.2	&	 +6, +10	&		&	99may	\\
269.153\(-\)1.128	&	09 03 33.46	&	 \(-\)48 28 02.6	&	1.6	&	 +16.0	&	 +7, +16	&		&	99may	\\
270.255+0.835	&	09 16 41.51	&	 \(-\)47 56 12.1	&	0.6	&	 +3.9	&	 +3, +5	&		&	99oct	\\
284.352\(-\)0.419	&	10 24 10.89	&	 \(-\)57 52 38.8	&	1.7	&	 +3.3	&	 +3, +11	&		&	C97	\\
285.337\(-\)0.002	&	10 32 09.64	&	 \(-\)58 02 05.2	&	10	&	 +0.5	&	 \(-\)8, +3	&		&	93nov	\\
287.371+0.644	&	10 48 04.40	&	 \(-\)58 27 01.7	&	80	&	\(-\)1.8	&	 \(-\)3, 0	&	c98	&	93nov	\\
290.374+1.661	&	11 12 18.10	&	 \(-\)58 46 21.5	&	0.6	&	\(-\)24.2	&	 \(-\)28, \(-\)22	&	c98	&	99feb	\\
291.274\(-\)0.709	&	11 11 53.35	&	 \(-\)61 18 23.7	&	100	&	\(-\)29.6	&	 \(-\)31, \(-\)28	&	c04	&	C04	\\
291.270\(-\)0.719	&	11 11 49.44	&	 \(-\)61 18 51.9	&	4	&	\(-\)26.5	&	 \(-\)32, \(-\)26	&		&	C04	\\
291.579\(-\)0.431	&	11 15 05.76	&	 \(-\)61 09 40.8	&	0.7	&	 +14.5	&	 +11, +16	&	c98	&	C04	\\
291.582\(-\)0.435	&	11 15 06.61	&	 \(-\)61 09 58.3	&	1.7	&	 +10.5	&	 +8, +11	&		&	C04	\\
294.511\(-\)1.621	&	11 35 32.25	&	 \(-\)63 14 43.2	&	12	&	\(-\)10.2	&	 \(-\)14, \(-\)9	&	c98	&	C97	\\
294.990\(-\)1.719	&	11 39 22.88	&	 \(-\)63 28 26.4	&	18	&	\(-\)12.3	&	 \(-\)13, \(-\)11	&		&	99oct	\\
296.893\(-\)1.305	&	11 56 50.07	&	 \(-\)63 32 05.5	&	2.5	&	 +22.2	&	 +21, +23	&		&	99oct	\\
298.213\(-\)0.343	&	12 09 55.18	&	 \(-\)62 50 01.1	&	1.4	&	 +37.2	&	  +33, +39	&		&	97may	\\
299.013+0.128	&	12 17 24.60	&	 \(-\)62 29 03.7	&	7	&	  +18.4	&	 +18, +20	&	c98	&	99may	\\
300.504\(-\)0.176	&	12 30 03.58	&	 \(-\)62 56 48.7	&	4.7	&	 +7.5	&	 +4, +11	&	c98	&	99may	\\
300.969+1.148	&	12 34 53.29	&	 \(-\)61 39 40.0	&	5.5	&	\(-\)37.2	&	 \(-\)40, \(-\)35	&	c98	&	C97	\\
301.136\(-\)0.226	&	12 35 35.14	&	 \(-\)63 02 32.6	&	1.2	&	 \(-\)39.8	&	 \(-\)41, \(-\)37	&	c98	&	99feb	\\
302.032\(-\)0.061	&	12 43 31.92	&	 \(-\)62 55 06.7	&	11	&	\(-\)35.3	&	 \(-\)43, \(-\)33	&		&	99oct, 00jun	\\
305.200+0.019	&	13 11 16.93	&	 \(-\)62 45 55.1	&	44	&	\(-\)33.1	&	 \(-\)38, \(-\)29	&	c98	&	C97	\\
305.199+0.005	&	13 11 17.20	&	 \(-\)62 46 46.0	&	2.3	&	\(-\)42.8	&	 \(-\)45, \(-\)38	&		&	C97	\\
305.202+0.208	&	13 11 10.49	&	 \(-\)62 34 38.8	&	20	&	 \(-\)43.9	&	 \(-\)47, \(-\)43	&	c98	&	CVF95	\\
305.208+0.206	&	13 11 13.71	&	 \(-\)62 34 41.4	&	320	&	\(-\)38.3	&	 \(-\)42, \(-\)34	&	c98	&	CVF95	\\
305.248+0.245	&	13 11 32.47	&	 \(-\)62 32 09.1	&	4	&	\(-\)32.0	&	 \(-\)36, \(-\)28	&		&	99oct	\\
305.362+0.150	&	13 12 35.86	&	 \(-\)62 37 17.9	&	3	&	 \(-\)36.5	&	 \(-\)38, \(-\)35	&	c98	&	99may	\\
305.366+0.184	&	13 12 36.74	&	 \(-\)62 35 14.7	&	2.5	&	\(-\)33.8	&	 \(-\)35, \(-\)33	&		&	99may	\\
305.799\(-\)0.245	&	13 16 43.23	&	 \(-\)62 58 32.9	&	0.7	&	 \(-\)39.5	&	 \(-\)40, \(-\)36	&	c98	&	99feb	\\
305.887+0.017	&	13 17 15.53	&	 \(-\)62 42 23.0	&	5.5	&	\(-\)34.0	&	 \(-\)35, \(-\)33	&		&	99oct	\\
306.322\(-\)0.334	&	13 21 23.01	&	 \(-\)63 00 29.5	&	1.2	&	 \(-\)24.4	&	 \(-\)25, \(-\)22	&	c98	&	99feb	\\
308.754+0.549	&	13 40 57.60	&	 \(-\)61 45 43.4	&	5	&	 \(-\)51.0	&	 \(-\)52, \(-\)39	&	c04	&	C04	\\
308.918+0.123	&	13 43 01.85	&	 \(-\)62 08 52.2	&	54	&	 \(-\)54.7	&	 \(-\)56, \(-\)52	&	c98	&	96dec	\\
309.384\(-\)0.135	&	13 47 23.98	&	 \(-\)62 18 12.0	&	1	&	 \(-\)49.6	&	 \(-\)51, \(-\)49	&	c98	&	99feb	\\
309.921+0.479	&	13 50 41.78	&	 \(-\)61 35 10.2	&	635	&	\(-\)59.8	&	 \(-\)65, \(-\)54	&	c98	&	C97	\\
310.144+0.760	&	13 51 58.43	&	 \(-\)61 15 41.3	&	120	&	 \(-\)55.6	&	 \(-\)59, \(-\)54	&	c98	&	96dec, 99oct	\\
311.643\(-\)0.380	&	14 06 38.77	&	 \(-\)61 58 23.1	&	11.6	&	32.5	&	 +31, +36	&	c98	&	C97	\\
311.947+0.142	&	14 07 49.72	&	 \(-\)61 23 08.3	&	0.6	&	 \(-\)38.3	&	 \(-\)39, \(-\)38	&	text	&	99feb	\\
312.108+0.262	&	14 08 49.31	&	 \(-\)61 13 25.1	&	15	&	\(-\)50.0	&	 \(-\)54, \(-\)49	&		&	99oct	\\
312.598+0.045	&	14 13 15.03	&	 \(-\)61 16 53.6	&	9	&	 \(-\)67.9	&	 \(-\)69, \(-\)64	&	c98	&	99feb	\\
312.597+0.045	&	14 13 14.35	&	 \(-\)61 16 57.7	&	1	&	 \(-\)60.0	&	 \(-\)61, \(-\)59	&		&	99feb	\\
313.469+0.190	&	14 19 40.94	&	 \(-\)60 51 47.3	&	16	&	 \(-\)9.4	&	 \(-\)13, \(-\)3	&	c98	&	99may	\\
313.577+0.325	&	14 20 08.58	&	 \(-\)60 42 00.8	&	70	&	 \(-\)47.9	&	 \(-\)54, \(-\)46	&	c98	&	96oct	\\
\hline
\end{tabular}
\medskip\\
\end{center}
\end{table*}

\begin{table*}[h]
\addtocounter{table}{-1}
\begin{center}
\caption{\textit{- continued p2 of 8}}\label{tableexample}
\begin{tabular}{lllrrccl}
\hline Galactic name & RA(2000) & Dec(2000) & I(pk) & v(pk) & 
v(range) & OH? & refs, epoch $^a$ \\
\ (~~l,~~~~~~~b) & (h~~m~~~s) & ~~$^\circ$~~~ '~~~~" & (Jy) & (\kms) 
& (\kms) & & \\
\hline
313.705\(-\)0.190	&	14 22 34.82	&	 \(-\)61 08 27.1	&	1.2	&	 \(-\)41.5	&	 \(-\)46, \(-\)41	&	c98	&	96dec	\\
313.767\(-\)0.863	&	14 25 01.73	&	 \(-\)61 44 58.1	&	15	&	 \(-\)54.6	&	 \(-\)57, \(-\)53	&	c98	&	96dec	\\
313.774\(-\)0.863	&	14 25 04.78	&	 \(-\)61 44 50.3	&	7	&	 \(-\)41.2	&	 \(-\)46, \(-\)40	&		&	96dec	\\
314.320+0.112	&	14 26 26.20	&	 \(-\)60 38 31.3	&	24	&	 \(-\)43.7	&	 \(-\)59, \(-\)43	&	c98	&	96oct	\\
316.359\(-\)0.362	&	14 43 11.20	&	 \(-\)60 17 13.3	&	52	&	 +3.5	&	 +1, +8	&	c98	&	96dec	\\
316.381\(-\)0.379	&	14 43 24.21	&	 \(-\)60 17 37.4	&	38	&	 \(-\)0.7	&	  \(-\)6, +1	&		&	96dec	\\
316.412\(-\)0.308	&	14 43 23.34	&	 \(-\)60 13 00.9	&	9	&	 \(-\)5.7	&	 \(-\)7, \(-\)2	&	c98	&	96dec	\\
316.640\(-\)0.087	&	14 44 18.45	&	 \(-\)59 55 11.5	&	60	&	 \(-\)19.8	&	 \(-\)25, \(-\)15	&	c98	&	96dec	\\
316.811\(-\)0.057	&	14 45 26.43	&	 \(-\)59 49 16.3	&	10	&	 \(-\)46.3	&	 \(-\)49, \(-\)42	&	c98	&	96dec	\\
317.701+0.110	&	14 51 11.69	&	 \(-\)59 17 02.1	&	9	&	 \(-\)43.6	&	 \(-\)46, \(-\)40	&		&	99oct	\\
318.043\(-\)1.404	&	14 59 08.61	&	 \(-\)60 28 25.5	&	6	&	+46.2	&	 +44, +47	&	c98	&	99oct	\\
318.050+0.087	&	14 53 42.67	&	 \(-\)59 08 52.4	&	10	&	 \(-\)46.5	&	 \(-\)59, \(-\)46	&	c98	&	99feb	\\
318.948\(-\)0.196	&	15 00 55.39	&	 \(-\)58 58 52.8	&	690	&	\(-\)34.7	&	 \(-\)39, \(-\)31	&	c98	&	CVF95	\\
319.836\(-\)0.197	&	15 06 54.65	&	 \(-\)58 33 00.0	&	0.4	&	\(-\)9.1	&	 \(-\)14, \(-\)9	&	c98	&	99may	\\
320.123\(-\)0.504	&	15 10 00.17	&	 \(-\)58 40 18.0	&	2.5	&	\(-\)10.1	&	 \(-\)12, \(-\)9	&		&	96dec	\\
320.231\(-\)0.284	&	15 09 51.94	&	 \(-\)58 25 38.5	&	21	&	\(-\)62.0	&	 \(-\)71, \(-\)58	&	c98	&	99may	\\
321.030\(-\)0.485	&	15 15 51.79	&	 \(-\)58 11 18.0	&	6	&	 \(-\)66.5	&	 \(-\)68, \(-\)56	&	c98	&	96dec	\\
321.033\(-\)0.483	&	15 15 52.63	&	 \(-\)58 11 07.7	&	50	&	 \(-\)61.6	&	 \(-\)69, \(-\)54	&		&	96dec	\\
321.148\(-\)0.529	&	15 16 48.39	&	 \(-\)58 09 50.2	&	7.5	&	\(-\)66.1	&	 \(-\)67, \(-\)65	&	c98	&	96dec	\\
322.158+0.636	&	15 18 34.64 	&	 \(-\)56 38 25.3	&	225	&	 \(-\)63.3	&	 \(-\)66, \(-\)51	&	c98	&	CVF95, 99oct	\\
323.459\(-\)0.079	&	15 29 19.33	&	 \(-\)56 31 22.8	&	19.2	&	\(-\)66.9	&	 \(-\)69, \(-\)66	&	c98	&	C97	\\
323.740\(-\)0.263	&	15 31 45.45	&	 \(-\)56 30 50.1	&	3000	&	\(-\)51.1	&	 \(-\)58, \(-\)45	&	c98	&	CVF95	\\
324.716+0.342	&	15 34 57.47	&	 \(-\)55 27 23.6	&	7	&	 \(-\)46.0	&	 \(-\)51, \(-\)45	&	c98	&	99feb	\\
326.475+0.703	&	15 43 16.64	&	 \(-\)54 07 14.6	&	60	&	 \(-\)38.3	&	 \(-\)51, \(-\)37	&		&	96dec, 00nov	\\
326.476+0.695	&	15 43 18.90	&	 \(-\)54 07 35.5	&	2.5	&	\(-\)43.6	&	 \(-\)44, \(-\)43	&		&	00nov	\\
326.641+0.611	&	15 44 33.33	&	 \(-\)54 05 31.5	&	16	&	\(-\)42.8	&	 \(-\)44, \(-\)36	&		&	96dec	\\
326.662+0.521	&	15 45 02.95	&	 \(-\)54 09 03.1	&	7	&	 \(-\)38.6	&	 \(-\)42, \(-\)38	&		&	96dec	\\
326.859\(-\)0.677	&	15 51 14.19	&	 \(-\)54 58 04.8	&	10	&	 \(-\)58.0	&	 \(-\)60, \(-\)57	&		&	00nov	\\
327.120+0.511	&	15 47 32.73	&	 \(-\)53 52 38.4	&	85	&	 \(-\)87.0	&	 \(-\)90, \(-\)83	&	c98	&	97may	\\
327.291\(-\)0.578	&	15 53 07.70	&	 \(-\)54 37 06.5	&	2.5	&	 \(-\)36.8	&	 \(-\)44, \(-\)36	&	c98	&	96dec	\\
327.392+0.199	&	15 50 18.48	&	 \(-\)53 57 06.3	&	7	&	\(-\)84.6	&	 \(-\)86, \(-\)79	&		&	00nov	\\
327.395+0.197	&	15 50 20.06	&	 \(-\)53 57 07.5	&	1.7	&	 \(-\)89.0	&	 \(-\)90, \(-\)88	&		&	00nov	\\
327.402+0.444	&	15 49 19.50	&	 \(-\)53 45 13.9	&	65	&	 \(-\)82.6	&	 \(-\)84, \(-\)72	&	c98	&	99may	\\
327.590\(-\)0.094	&	15 52 36.82	&	 \(-\)54 03 18.7	&	3.2	&	\(-\)86.2	&	 \(-\)87, \(-\)85	&		&	00nov	\\
327.618\(-\)0.111	&	15 52 50.22	&	 \(-\)54 03 00.5	&	3.1	&	\(-\)97.6	&	 \(-\)99, \(-\)97	&		&	00nov	\\
327.945\(-\)0.115	&	15 54 33.91	&	 \(-\)53 50 44.3	&	6.9	&	\(-\)51.6	&	 \(-\)52, \(-\)51	&		&	00nov	\\
328.237\(-\)0.547	&	15 57 58.31	&	 \(-\)53 59 23.0	&	360	&	\(-\)44.5	&	 \(-\)47, \(-\)31	&	c98	&	CVF95	\\
328.254\(-\)0.532	&	15 57 59.78	&	 \(-\)53 58 00.8	&	440	&	\(-\)37.5	&	 \(-\)51, \(-\)36	&	c98	&	CVF95	\\
328.808+0.633	&	15 55 48.45	&	 \(-\)52 43 06.6	&	240	&	\(-\)43.8	&	 \(-\)47, \(-\)42	&	c98	&	C97	\\
328.809+0.633	&	15 55 48.70	&	 \(-\)52 43 05.5	&	53	&	\(-\)44.2	&	 \(-\)45, \(-\)43	&	c98	&	C97	\\
329.029\(-\)0.205	&	16 00 31.80	&	 \(-\)53 12 49.6	&	138	&	\(-\)37.4	&	 \(-\)42, \(-\)34	&	c98	&	CVF95	\\
329.031\(-\)0.198	&	16 00 30.32	&	 \(-\)53 12 27.3	&	11	&	\(-\)45.5	&	 \(-\)49, \(-\)42	&	c98	&	CVF95	\\
329.066\(-\)0.308	&	16 01 09.93	&	 \(-\)53 16 02.6	&	20	&	 \(-\)43.8	&	 \(-\)46, \(-\)43	&	c98	&	CVF95	\\
329.183\(-\)0.314	&	16 01 47.01	&	 \(-\)53 11 43.3	&	10	&	 \(-\)55.7	&	 \(-\)60, \(-\)50	&	c98	&	99feb	\\
329.339+0.148	&	16 00 33.13	&	 \(-\)52 44 39.8	&	18	&	 \(-\)106.6	&	 \(-\)108, \(-\)105	&	c01	&	00nov	\\
329.405\(-\)0.459	&	16 03 32.16	&	 \(-\)53 09 30.5	&	33	&	 \(-\)70.5	&	 \(-\)73, \(-\)63	&	c98	&	98nov	\\
329.407\(-\)0.459	&	16 03 32.65	&	 \(-\)53 09 26.9	&	72	&	 \(-\)66.7	&	 \(-\)68, \(-\)66	&		&	98nov	\\
329.469+0.502	&	15 59 40.76	&	 \(-\)52 23 27.7	&	8	&	\(-\)72.0	&	 \(-\)74, \(-\)65	&		&	99oct	\\
329.610+0.114	&	16 02 03.14	&	 \(-\)52 35 33.5	&	30	&	\(-\)60.0	&	 \(-\)69, \(-\)59	&		&	99oct	\\
329.622+0.138	&	16 02 00.33	&	 \(-\)52 33 59.4	&	1.9	&	\(-\)84.8	&	 \(-\)86, \(-\)83	&		&	99oct	\\
\hline
\end{tabular}
\medskip\\
\end{center}
\end{table*}

\begin{table*}[h]
\addtocounter{table}{-1}
\begin{center}
\caption{\textit{- continued p3 of 8}}\label{tableexample}
\begin{tabular}{lllrrccl}
\hline Galactic name & RA(2000) & Dec(2000) & I(pk) & v(pk) &   
v(range) & OH? & refs, epoch $^a$ \\
\ (~~l,~~~~~~~b) & (h~~m~~~s) & ~~$^\circ$~~~ '~~~~" & 
(Jy) & (\kms)
& (\kms) & & \\
\hline
330.070+1.064	&	16 00 15.43	&	 \(-\)51 34 25.6	&	8	&	 \(-\)38.8	&	 \(-\)56, \(-\)37	&		&	99oct	\\
330.878\(-\)0.367	&	16 10 19.79	&	 \(-\)52 06 07.8	&	0.6	&	\(-\)59.3	&	 \(-\)60, \(-\)58	&	c98	&	97may	\\
330.875\(-\)0.383	&	16 10 23.09	&	 \(-\)52 06 58.7	&	0.4	&	 \(-\)56.5	&	  \(-\)72, \(-\)56	&		&	97may	\\
330.953\(-\)0.182	&	16 09 52.37	&	 \(-\)51 54 57.6	&	7	&	\(-\)87.6	&	 \(-\)90, \(-\)87	&	c01	&	97may	\\
331.120\(-\)0.118	&	16 10 23.05	&	 \(-\)51 45 20.1	&	9.4	&	\(-\)93.2	&	 \(-\)95, \(-\)90	&		&	C96a	\\
331.132\(-\)0.244	&	16 10 59.76	&	 \(-\)51 50 22.6	&	40	&	 \(-\)84.3	&	 \(-\)92, \(-\)81	&	c98	&	97may	\\
331.278\(-\)0.188	&	16 11 26.59	&	 \(-\)51 41 56.7	&	190	&	\(-\)78.2	&	 \(-\)87. \(-\)77	&	c98	&	CVF95, C96a	\\
331.342\(-\)0.346	&	16 12 26.45	&	 \(-\)51 46 16.4	&	56	&	 \(-\)67.4	&	 \(-\)70, \(-\)62	&	c98	&	99may	\\
331.425+0.264	&	16 10 09.33	&	 \(-\)51 16 04.3	&	18	&	\(-\)88.7	&	 \(-\)91, \(-\)78	&		&	00nov	\\
331.442\(-\)0.187	&	16 12 12.49	&	 \(-\)51 35 10.1	&	81	&	\(-\)88.4	&	 \(-\)93, \(-\)84	&	c98	&	C96a	\\
331.542\(-\)0.066	&	16 12 09.02	&	 \(-\)51 25 47.6	&	7	&	\(-\)85.9	&	 \(-\)87, \(-\)85	&	c98	&	C96a; C97	\\
331.543\(-\)0.066	&	16 12 09.14	&	 \(-\)51 25 45.3	&	11.6	&	\(-\)84.1	&	 \(-\)85, \(-\)83	&	c98	&	C96a; C97	\\
331.556\(-\)0.121	&	16 12 27.21	&	 \(-\)51 27 38.2	&	35	&	\(-\)103.4	&	 \(-\)105, \(-\)96	&	c98	&	C96a, C97	\\
332.094\(-\)0.421	&	16 16 16.45	&	 \(-\)51 18 25.7	&	11	&	 \(-\)58.6	&	 \(-\)62, \(-\)58	&		&	00nov	\\
332.295+2.280	&	16 05 41.72	&	 \(-\)49 11 30.3	&	113	&	\(-\)24.0	&	 \(-\)27, \(-\)20	&	c98	&	97may	\\
332.295\(-\)0.094	&	16 15 45.38	&	 \(-\)50 55 53.4	&	6.3	&	\(-\)47.0	&	 \(-\)48, \(-\)42	&		&	95jul	\\
332.351\(-\)0.436	&	16 17 31.51	&	 \(-\)51 08 22.0	&	2.6	&	\(-\)53.2	&	 \(-\)55, \(-\)52	&		&	00nov	\\
332.352\(-\)0.117	&	16 16 07.08	&	 \(-\)50 54 31.0	&	1	&	 \(-\)41.8	&	 \(-\)43, \(-\)41	&	c98	&	94may, 95jul	\\
332.560\(-\)0.148	&	16 17 12.11	&	 \(-\)50 47 12.3	&	5.1	&	\(-\)55.6	&	 \(-\)56, \(-\)55	&		&	C96a	\\
332.604\(-\)0.167	&	16 17 29.31	&	 \(-\)50 46 12.5	&	6.8	&	\(-\)50.9	&	 \(-\)52, \(-\)50	&		&	C96a	\\
332.653\(-\)0.621	&	16 19 43.51	&	 \(-\)51 03 36.9	&	7.1	&	 \(-\)50.6	&	 \(-\)52, \(-\)49	&		&	99may	\\
332.701\(-\)0.587	&	16 19 47.42	&	 \(-\)51 00 09.5	&	2.1	&	\(-\)62.7	&	 \(-\)63, \(-\)62	&		&	99may	\\
332.726\(-\)0.621	&	16 20 03.00	&	 \(-\)51 00 32.5	&	2.7	&	 \(-\)49.5	&	 \(-\)57, \(-\)44	&	c98	&	99may	\\
332.826\(-\)0.549	&	16 20 10.85	&	 \(-\)50 53 14.1	&	1.6	&	 \(-\)61.7	&	 \(-\)62, \(-\)55	&		&	97may	\\
332.942\(-\)0.686	&	16 21 19.00	&	 \(-\)50 54 10.2	&	10.3	&	\(-\)52.8	&	 \(-\)55, \(-\)52	&		&	00nov	\\
332.963\(-\)0.679	&	16 21 22.92	&	 \(-\)50 52 58.5	&	35	&	\(-\)45.8	&	 \(-\)49, \(-\)45	&		&	00nov	\\
333.029\(-\)0.015	&	16 18 44.18	&	 \(-\)50 21 50.6	&	3.6	&	\(-\)55.2	&	 \(-\)62, \(-\)52	&		&	C96a	\\
333.029\(-\)0.063	&	16 18 56.73	&	 \(-\)50 23 54.1	&	3.9	&	\(-\)40.4	&	 \(-\)41, \(-\)40	&		&	C96a	\\
333.068\(-\)0.447	&	16 20 48.95	&	 \(-\)50 38 40.2	&	14.1	&	\(-\)54.5	&	 \(-\)57, \(-\)51	&		&	C97	\\
333.121\(-\)0.434	&	16 20 59.66	&	 \(-\)50 35 51.9	&	18.9	&	\(-\)49.3	&	 \(-\)51, \(-\)48	&		&	C97	\\
333.126\(-\)0.440	&	16 21 02.61	&	 \(-\)50 35 54.7	&	3.9	&	\(-\)43.9	&	 \(-\)44, \(-\)41	&		&	C97	\\
333.128\(-\)0.440	&	16 21 03.26	&	 \(-\)50 35 49.4	&	3.6	&	\(-\)44.6	&	 \(-\)47, \(-\)44	&		&	C97	\\
333.128\(-\)0.560	&	16 21 35.38	&	 \(-\)50 40 56.5	&	14.5	&	 \(-\)52.7	&	 \(-\)61, \(-\)52	&		&	00nov	\\
333.130\(-\)0.560	&	16 21 35.73	&	 \(-\)50 40 51.0	&	17	&	\(-\)56.8	&	 \(-\)64, \(-\)56	&		&	00nov	\\
333.135\(-\)0.431s	&	16 21 02.82	&	 \(-\)50 35 12.0	&	1	&	\(-\)53.0	&	 \(-\)54, \(-\)51	&	c98	&	C97	\\
333.163\(-\)0.101	&	16 19 42.67	&	 \(-\)50 19 53.2	&	7.7	&	\(-\)95.3	&	 \(-\)96. \(-\)90	&		&	C96a	\\
333.184\(-\)0.091	&	16 19 45.62	&	 \(-\)50 18 35.0	&	7.1	&	\(-\)84.7	&	 \(-\)91, \(-\)81	&		&	C96a	\\
333.234\(-\)0.062	&	16 19 51.25	&	 \(-\)50 15 14.1	&	1.9	&	\(-\)91.9	&	 \(-\)93, \(-\)79	&		&	C96a	\\
333.315+0.105	&	16 19 29.01	&	 \(-\)50 04 41.3	&	9.6	&	\(-\)45.0	&	 \(-\)51, \(-\)40	&	c98	&	C96a	\\
333.387+0.032	&	16 20 07.59	&	 \(-\)50 04 46.5	&	3.4	&	 \(-\)73.9	&	 \(-\)75, \(-\)60	&	c98	&	95jul	\\
333.466\(-\)0.164	&	16 21 20.18	&	 \(-\)50 09 48.6	&	63	&	\(-\)42.5	&	 \(-\)49, \(-\)37	&	c98	&	C96a	\\
333.562\(-\)0.025	&	16 21 08.80	&	 \(-\)49 59 48.0	&	46.7	&	\(-\)35.8	&	 \(-\)45, \(-\)33	&		&	C96a	\\
333.646+0.058	&	16 21 09.14	&	 \(-\)49 52 45.9	&	2.9	&	\(-\)87.3	&	 \(-\)89, \(-\)82	&		&	C96a	\\
333.683\(-\)0.437	&	16 23 29.78	&	 \(-\)50 12 08.6	&	20	&	\(-\)5.3	&	 \(-\)8, 0	&		&	00nov	\\
333.931\(-\)0.135	&	16 23 14.83	&	 \(-\)49 48 48.9	&	10.9	&	\(-\)36.7	&	 \(-\)39, \(-\)36	&		&	C96a	\\
334.635\(-\)0.015	&	16 25 45.73	&	 \(-\)49 13 37.4	&	36.4	&	\(-\)30.0	&	 \(-\)31, \(-\)27	&		&	C96a	\\
334.935\(-\)0.098	&	16 27 24.25	&	 \(-\)49 04 11.3	&	7.1	&	\(-\)19.5	&	 \(-\)22, \(-\)17	&		&	C96a	\\
335.060\(-\)0.427	&	16 29 23.13	&	 \(-\)49 12 27.1	&	20	&	\(-\)47.0	&	 \(-\)48, \(-\)25	&	c98	&	97may	\\
335.556\(-\)0.307	&	16 30 55.98	&	 \(-\)48 45 50.2	&	23	&	\(-\)116.4	&	 \(-\)119, \(-\)110	&	c98	&	96oct	\\
335.585\(-\)0.285	&	16 30 57.28	&	 \(-\)48 43 39.7	&	19.8	&	\(-\)49.3	&	 \(-\)51, \(-\)43	&	c98	&	CVF95, C96a	\\
\hline
\end{tabular}
\medskip\\
\end{center}
\end{table*}

\begin{table*}[h]
\addtocounter{table}{-1}
\begin{center}
\caption{\textit{- continued p4 of 8}}\label{tableexample}
\begin{tabular}{lllrrccl}
\hline Galactic name & RA(2000) & Dec(2000) & I(pk) & v(pk) &
v(range) & OH? & refs, epoch $^a$ \\
\ (~~l,~~~~~~~b) & (h~~m~~~s) & ~~$^\circ$~~~ '~~~~" & (Jy) & 
(\kms)
& (\kms) & & \\
\hline
335.585\(-\)0.289	&	16 30 58.67	&	 \(-\)48 43 50.7	&	31	&	 \(-\)51.4	&	 \(-\)56, \(-\)50	&	c98	&	CVF95, C96a	\\
335.585\(-\)0.290	&	16 30 58.79	&	 \(-\)48 43 53.4	&	108	&	 \(-\)47.3	&	 \(-\)48, \(-\)45	&		&	CVF95, C96a	\\
335.726+0.191	&	16 29 27.37	&	 \(-\)48 17 53.2	&	78	&	\(-\)44.4	&	 \(-\)55, \(-\)43	&		&	CVF95, C96a	\\
335.789+0.174	&	16 29 47.33	&	 \(-\)48 15 51.7	&	118	&	\(-\)47.6	&	 \(-\)59, \(-\)45	&	c98	&	CVF95, C06a	\\
336.018\(-\)0.827	&	16 35 09.26	&	 \(-\)48 46 47.4	&	300	&	 \(-\)53.4	&	 \(-\)55, \(-\)39	&	c98	&	97may	\\
336.358\(-\)0.137	&	16 33 29.17	&	 \(-\)48 03 43.9	&	18.3	&	\(-\)73.6	&	 \(-\)82, \(-\)72	&	c98	&	C96a	\\
336.410\(-\)0.258	&	16 34 13.20	&	 \(-\)48 06 20.9	&	8	&	\(-\)85.6	&	 \(-\)87, \(-\)84	&		&	96dec	\\
336.433\(-\)0.262	&	16 34 20.22	&	 \(-\)48 05 32.2	&	46	&	\(-\)93.3	&	 \(-\)95, \(-\)86	&		&	96dec	\\
336.496\(-\)0.271	&	16 34 38.02	&	 \(-\)48 03 03.9	&	11	&	 \(-\)23.9	&	 \(-\)25, \(-\)19	&		&	96dec	\\
336.822+0.028	&	16 34 38.28	&	 \(-\)47 36 32.2	&	19.5	&	\(-\)76.7	&	 \(-\)78, \(-\)76	&	c98	&	C96a	\\
336.830\(-\)0.375	&	16 36 26.19	&	 \(-\)47 52 31.1	&	36	&	 \(-\)22.7	&	  \(-\)28, \(-\)21	&		&	99oct	\\
336.864+0.005	&	16 34 54.44	&	 \(-\)47 35 37.3	&	32	&	\(-\)76.1	&	 \(-\)83, \(-\)73	&	c98	&	C96a	\\
336.941\(-\)0.156	&	16 35 55.19	&	 \(-\)47 38 45.4	&	33.9	&	\(-\)67.3	&	 \(-\)79, \(-\)64	&	c98	&	C96a	\\
336.983\(-\)0.183	&	16 36 12.29	&	 \(-\)47 37 56.2	&	18.8	&	\(-\)80.8	&	 \(-\)82, \(-\)78	&	c98	&	C96a	\\
336.994\(-\)0.027	&	16 35 33.98	&	 \(-\)47 31 11.7	&	28.2	&	\(-\)125.8	&	 \(-\)127, \(-\)118	&	c98	&	C96a	\\
337.176\(-\)0.032	&	16 36 18.84	&	 \(-\)47 23 19.8	&	9.4	&	\(-\)64.7	&	 \(-\)74, \(-\)64	&		&	C96a	\\
337.258\(-\)0.101	&	16 36 56.32	&	 \(-\)47 22 26.4	&	15	&	\(-\)69.3	&	 \(-\)74, \(-\)67	&	c98	&	C96a	\\
337.404\(-\)0.402	&	16 38 50.52	&	 \(-\)47 28 00.2	&	67	&	 \(-\)39.7	&	 \(-\)43, \(-\)37	&	c98	&	97may	\\
337.613\(-\)0.060	&	16 38 09.54	&	 \(-\)47 04 59.9	&	20	&	 \(-\)42.0	&	 \(-\)54, \(-\)38	&	c98	&	C96a, C97	\\
337.632\(-\)0.079	&	16 38 19.12	&	 \(-\)47 04 53.3	&	13.7	&	\(-\)56.9	&	 \(-\)64, \(-\)54	&		&	C96a, C97	\\
337.686+0.137	&	16 37 35.42	&	 \(-\)46 53 47.6	&	2.1	&	\(-\)74.9	&	 \(-\)76, \(-\)74	&		&	C96a	\\
337.703\(-\)0.053	&	16 38 29.12	&	 \(-\)47 00 43.2	&	6	&	\(-\)44	&	 \(-\)52, \(-\)43	&		&	C96a	\\
337.705\(-\)0.053	&	16 38 29.63	&	 \(-\)47 00 35.5	&	145	&	\(-\)54.6	&	 \(-\)57, \(-\)49	&	c98	&	CVF95, C96a	\\
337.710+0.089	&	16 37 53.41	&	 \(-\)46 54 40.8	&	5	&	\(-\)72.6	&	 \(-\)74, \(-\)72	&		&	C96a	\\
337.920\(-\)0.456	&	16 41 06.05	&	 \(-\)47 07 02.5	&	28	&	 \(-\)38.8	&	 \(-\)41, \(-\)36	&	c98	&	00jun	\\
337.966\(-\)0.169	&	16 40 01.09	&	 \(-\)46 53 34.5	&	10.6	&	\(-\)60.4	&	 \(-\)61, \(-\)54	&		&	C96a	\\
337.997+0.136	&	16 38 48.50	&	 \(-\)46 39 57.5	&	4	&	\(-\)32.3	&	 \(-\)36, \(-\)31	&	c98	&	96dec	\\
338.075+0.012	&	16 39 39.07	&	 \(-\)46 41 28.1	&	18.8	&	\(-\)53.0	&	 \(-\)55, \(-\)43	&	c98	&	C96a	\\
338.075+0.009	&	16 39 39.81	&	 \(-\)46 41 33.1	&	3.5	&	\(-\)38.2	&	 \(-\)42, \(-\)35	&		&	C96a	\\
338.280+0.542	&	16 38 09.08	&	 \(-\)46 11 03.3	&	4.8	&	\(-\)56.8	&	 \(-\)59, \(-\)56	&	c98	&	C97	\\
338.287+0.120	&	16 40 00.13	&	 \(-\)46 27 37.1	&	11.9	&	\(-\)39.3	&	 \(-\)44, \(-\)38	&		&	C96a	\\
338.432+0.058	&	16 40 49.79	&	 \(-\)46 23 37.0	&	49.6	&	\(-\)30.2	&	 \(-\)34, \(-\)27	&		&	C96a	\\
338.461\(-\)0.245	&	16 42 15.50	&	 \(-\)46 34 18.4	&	74	&	\(-\)50.4	&	 \(-\)57, \(-\)49	&	c98	&	C96a	\\
338.472+0.289	&	16 39 58.91	&	 \(-\)46 12 35.4	&	0.7	&	\(-\)30.5	&	 \(-\)35, \(-\)29	&	c98	&	99feb	\\
338.561+0.218	&	16 40 37.96	&	 \(-\)46 11 25.8	&	35	&	\(-\)40.8	&	 \(-\)43, \(-\)37	&		&	99feb	\\
338.566+0.110	&	16 41 07.03	&	 \(-\)46 15 28.3	&	6.1	&	\(-\)75.0	&	 \(-\)80, \(-\)73	&		&	C96a	\\
338.875\(-\)0.084	&	16 43 08.25	&	 \(-\)46 09 12.8	&	19.4	&	\(-\)41.4	&	 \(-\)42, \(-\)39	&	c98	&	C96a	\\
338.925+0.557	&	16 40 33.53	&	 \(-\)45 41 37.1	&	5	&	 \(-\)62.3	&	 \(-\)66, \(-\)59	&	c98	&	97may	\\
338.920+0.550	&	16 40 34.01	&	 \(-\)45 42 07.1	&	55	&	 \(-\)61.4	&	 \(-\)68, \(-\)59	&		&	97may	\\
338.935\(-\)0.062	&	16 43 16.01	&	 \(-\)46 05 40.2	&	22.6	&	\(-\)41.9	&	 \(-\)44, \(-\)41	&		&	C96a	\\
339.053\(-\)0.315	&	16 44 48.99	&	 \(-\)46 10 13.0	&	141	&	\(-\)111.6	&	 \(-\)114, \(-\)111	&	c98	&	C96a	\\
339.064+0.152	&	16 42 49.56	&	 \(-\)45 51 23.8	&	6.9	&	\(-\)85.6	&	 \(-\)90, \(-\)83	&		&	C96a	\\
339.282+0.136	&	16 43 43.11	&	 \(-\)45 42 08.0	&	8.8	&	\(-\)69.1	&	 \(-\)72, \(-\)68	&	c98	&	C96a	\\
339.294+0.139	&	16 43 44.95	&	 \(-\)45 41 28.0	&	5	&	\(-\)74.6	&	 \(-\)76, \(-\)65	&		&	C96a	\\
339.477+0.043	&	16 44 50.98	&	 \(-\)45 36 56.1	&	2	&	\(-\)9.3	&	 \(-\)11, \(-\)5	&		&	C96a	\\
339.582\(-\)0.127	&	16 45 58.82	&	 \(-\)45 38 47.2	&	8.7	&	\(-\)31.3	&	 \(-\)32, \(-\)29	&		&	C96a	\\
339.622\(-\)0.121	&	16 46 05.99	&	 \(-\)45 36 43.3	&	95	&	\(-\)32.8	&	 \(-\)39, \(-\)32	&	c98	&	C96a	\\
339.681\(-\)1.208	&	16 51 06.21	&	 \(-\)46 16 02.9	&	41	&	\(-\)21.5	&	 \(-\)39, \(-\)20	&		&	CVF95, 97may	\\
339.682\(-\)1.207	&	16 51 06.23	&	 \(-\)46 15 58.1	&	5.8	&	 \(-\)34.0	&	 \(-\)35, \(-\)33	&	c98	&	97may	\\
339.762+0.054	&	16 45 51.56	&	 \(-\)45 23 32.6	&	12.9	&	\(-\)51.0	&	 \(-\)52, \(-\)49	&		&	C96a	\\
\hline
\end{tabular}
\medskip\\
\end{center}
\end{table*}

\begin{table*}[h]
\addtocounter{table}{-1}
\begin{center}
\caption{\textit{- continued p5 of 8}}\label{tableexample}
\begin{tabular}{lllrrccl}
\hline Galactic name & RA(2000) & Dec(2000) & I(pk) & v(pk) &
v(range) & OH? & refs, epoch $^a$ \\
\ (~~l,~~~~~~~b) & (h~~m~~~s) & ~~$^\circ$~~~ '~~~~" & (Jy) &
(\kms)
& (\kms) & & \\
\hline
339.884\(-\)1.259	&	16 52 04.66	&	 \(-\)46 08 34.2	&	1650	&	\(-\)38.7	&	 \(-\)41, \(-\)27	&	c98	&	CVF95, 97may	\\
340.054\(-\)0.244	&	16 48 13.89	&	 \(-\)45 21 43.5	&	40	&	 \(-\)59.7	&	 \(-\)63, \(-\)46	&	c98	&	97may	\\
340.518\(-\)0.152	&	16 49 31.36	&	 \(-\)44 56 54.6	&	6.4	&	 \(-\)48.2	&	 \(-\)51, \(-\)43	&		&	00nov	\\
340.785\(-\)0.096	&	16 50 14.84	&	 \(-\)44 42 26.3	&	144	&	\(-\)105.1	&	 \(-\)110, \(-\)86	&	c98	&	C97	\\
341.218\(-\)0.212	&	16 52 17.84	&	 \(-\)44 26 52.1	&	137	&	\(-\)37.9	&	 \(-\)50, \(-\)35	&	c98	&	CVF95	\\
341.276+0.062	&	16 51 19.41	&	 \(-\)44 13 44.5	&	4	&	 \(-\)73.8	&	 \(-\)77, \(-\)66	&	c98	&	99feb	\\
342.484+0.183	&	16 55 02.30	&	 \(-\)43 12 59.8	&	101	&	 \(-\)41.8	&	 \(-\)45, \(-\)38	&		&	00nov	\\
343.929+0.125	&	17 00 10.91	&	 \(-\)42 07 19.3	&	12	&	14.3	&	 +9, +19	&	c98	&	C97	\\
344.227\(-\)0.569	&	17 04 07.78	&	 \(-\)42 18 39.5	&	90	&	 \(-\)19.8	&	 \(-\)33, \(-\)16	&	c98	&	CVF95	\\
344.419+0.044	&	17 02 08.62	&	 \(-\)41 47 10.3	&	1.5	&	\(-\)63.5	&	 \(-\)66, \(-\)63	&	c98	&	98nov	\\
344.421+0.045	&	17 02 08.77	&	 \(-\)41 46 58.5	&	14	&	 \(-\)71.5	&	 \(-\)73, \(-\)70	&		&	98nov	\\
344.581\(-\)0.024	&	17 02 57.71	&	 \(-\)41 41 53.8	&	3	&	 +1.4	&	 \(-\)6, +3	&	c98	&	99feb	\\
345.003\(-\)0.223	&	17 05 10.89	&	 \(-\)41 29 06.2	&	240	&	\(-\)22.5	&	 \(-\)25, \(-\)20	&		&	C97	\\
345.003\(-\)0.224	&	17 05 11.23	&	 \(-\)41 29 06.9	&	73	&	\(-\)26.2	&	 \(-\)34, \(-\)25	&	c98	&	C97	\\
345.010+1.792	&	16 56 47.58	&	 \(-\)40 14 25.8	&	410	&	\(-\)18	&	 \(-\)24, \(-\)16	&	c98	&	C97	\\
345.012+1.797	&	16 56 46.82	&	 \(-\)40 14 08.9	&	31	&	\(-\)12.7	&	 \(-\)16, \(-\)10	&		&	C97	\\
345.407\(-\)0.952	&	17 09 35.42	&	 \(-\)41 35 57.1	&	1	&	\(-\)14.4	&	 \(-\)15, \(-\)14	&	c98	&	98nov	\\
345.424\(-\)0.951	&	17 09 38.56	&	 \(-\)41 35 04.6	&	1.8	&	\(-\)13.5	&	 \(-\)19, \(-\)13	&		&	98nov	\\
345.498+1.467	&	16 59 42.84	&	 \(-\)40 03 36.1	&	2.4	&	 \(-\)14.2	&	 \(-\)15, \(-\)13	&	c98	&	97may	\\
345.505+0.348	&	17 04 22.91	&	 \(-\)40 44 21.7	&	130	&	\(-\)17.7	&	 \(-\)23, \(-\)11	&	c98	&	CVF95, 99oct	\\
345.487+0.314	&	17 04 28.24	&	 \(-\)40 46 28.7	&	1.3	&	\(-\)22.6	&	 \(-\)24, \(-\)22	&		&	99oct	\\
346.481+0.132	&	17 08 22.72	&	 \(-\)40 05 25.6	&	1.9	&	 \(-\)5.5	&	 \(-\)12, \(-\)5	&	c98	&	99may. 99oct	\\
346.480+0.221	&	17 08 00.11	&	 \(-\)40 02 15.9	&	30	&	 \(-\)18.9	&	 \(-\)20, \(-\)14	&		&	99feb. 00nov	\\
346.517+0.117	&	17 08 33.20	&	 \(-\)40 04 14.3	&	1	&	\(-\)0.1	&	 \(-\)3, +1	&		&	99oct	\\
346.522+0.085	&	17 08 42.29	&	 \(-\)40 05 07.8	&	0.6	&	+5.5	&	 +5, +6	&		&	99oct	\\
347.583+0.213	&	17 11 26.72	&	 \(-\)39 09 22.5	&	2.4	&	\(-\)102.5	&	 \(-\)103, \(-\)96	&		&	C97	\\
347.628+0.149	&	17 11 50.92	&	 \(-\)39 09 29.2	&	13.5	&	\(-\)96.6	&	 \(-\)98, \(-\)95	&	c98	&	C97	\\
347.631+0.211	&	17 11 36.05	&	 \(-\)39 07 07.0	&	11.2	&	\(-\)91.9	&	 \(-\)94, \(-\)89	&		&	C97	\\
347.817+0.018	&	17 12 58.05	&	 \(-\)39 04 56.1	&	3.4	&	\(-\)25.6	&	 \(-\)26, \(-\)23	&		&	96dec	\\
347.863+0.019	&	17 13 06.23	&	 \(-\)39 02 40.0	&	7.2	&	 \(-\)29.3	&	 \(-\)38, \(-\)28	&		&	96dec	\\
347.902+0.052	&	17 13 05.11	&	 \(-\)38 59 35.5	&	5.3	&	 \(-\)27.8	&	 \(-\)31, \(-\)27	&		&	96dec	\\
348.550\(-\)0.979	&	17 19 20.41	&	 \(-\)39 03 51.6	&	37	&	\(-\)10.0	&	 \(-\)19, \(-\)7	&	c98	&	CVF95, C97	\\
348.550\(-\)0.979n	&	17 19 20.45	&	 \(-\)39 03 49.4	&	32	&	\(-\)20.0	&	 \(-\)23, \(-\)14	&		&	CVF95, C97	\\
348.579\(-\)0.920	&	17 19 10.61	&	 \(-\)39 00 24.2	&	0.5	&	\(-\)15	&	 \(-\)16, \(-\)14	&	c98	&	96oct	\\
348.703\(-\)1.043	&	17 20 04.06	&	 \(-\)38 58 30.9	&	60	&	 \(-\)3.3	&	 \(-\)17, \(-\)3	&	c98	&	96dec	\\
348.727\(-\)1.037	&	17 20 06.54	&	 \(-\)38 57 09.1	&	90	&	 \(-\)7.6	&	 \(-\)12, \(-\)6	&	c98	&	96dec	\\
348.884+0.096	&	17 15 50.13	&	 \(-\)38 10 12.4	&	5	&	 \(-\)76.2	&	 \(-\)77, \(-\)73	&	c98	&	99feb	\\
348.892\(-\)0.180	&	17 17 00.23	&	 \(-\)38 19 28.9	&	2	&	 +1.4	&	 +1, +2	&	c98	&	99oct, 00jun	\\
349.067\(-\)0.017	&	17 16 50.74	&	 \(-\)38 05 14.3	&	1.9	&	 +6.9	&	 +6, +16	&	c98	&	99feb	\\
349.092+0.105	&	17 16 24.74	&	 \(-\)37 59 47.2	&	30	&	 \(-\)76.5	&	 \(-\)78, \(-\)74	&		&	96dec	\\
349.092+0.106	&	17 16 24.59	&	 \(-\)37 59 45.8	&	6	&	\(-\)80.4	&	 \(-\)83, \(-\)78	&	c98	&	96dec	\\
350.011\(-\)1.342	&	17 25 06.54	&	 \(-\)38 04 00.7	&	0.4	&	 \(-\)25.8	&	 \(-\)28, \(-\)25	&	c98	&	99feb, 99may	\\
350.015+0.433	&	17 17 45.45	&	 \(-\)37 03 11.9	&	3.2	&	 \(-\)31.7	&	 \(-\)37, \(-\)29	&	c98	&	99may	\\
350.105+0.083	&	17 19 27.01	&	 \(-\)37 10 53.3	&	16	&	 \(-\)74.0	&	 \(-\)76, \(-\)61	&		&	96oct	\\
350.104+0.084	&	17 19 26.68	&	 \(-\)37 10 53.1	&	2.5	&	\(-\)68.4	&	 \(-\)69, \(-\)68	&		&	96oct	\\
350.116+0.084	&	17 19 28.83	&	 \(-\)37 10 18.8	&	1.8	&	\(-\)68.0	&	 \(-\)69, \(-\)67	&		&	96oct	\\
350.299+0.122	&	17 19 50.87	&	 \(-\)36 59 59.9	&	26	&	\(-\)62.1	&	 \(-\)67, \(-\)61	&		&	96oct	\\
350.344+0.116	&	17 20 00.03	&	 \(-\)36 58 00.1	&	17	&	\(-\)65.5	&	 \(-\)66, \(-\)59	&		&	96dec	\\
350.686\(-\)0.491	&	17 23 28.63	&	 \(-\)37 01 48.8	&	19	&	 \(-\)13.8	&	 \(-\)15, \(-\)13	&	c98	&	97may	\\
351.160+0.697	&	17 19 57.50	&	 \(-\)35 57 52.8	&	9	&	\(-\)5.2	&	 \(-\)7, \(-\)2	&	c98	&	99oct, 00jun	\\
\hline
\end{tabular}
\medskip\\
\end{center}
\end{table*}

\begin{table*}[h]
\addtocounter{table}{-1}
\begin{center}
\caption{\textit{- continued p6 of 8}}\label{tableexample}
\begin{tabular}{lllrrccl}
\hline Galactic name & RA(2000) & Dec(2000) & I(pk) & v(pk) &
v(range) & OH? & refs, epoch $^a$ \\
\ (~~l,~~~~~~~b) & (h~~m~~~s) & ~~$^\circ$~~~ '~~~~" & (Jy) &
(\kms)
& (\kms) & & \\
\hline
351.417+0.645	&	17 20 53.37	&	 \(-\)35 47 01.2	&	2500	&	\(-\)10.4	&	 \(-\)12, \(-\)6	&	c98	&	C97	\\
351.417+0.646	&	17 20 53.18	&	 \(-\)35 46 59.3	&	1600	&	\(-\)11.2	&	 \(-\)12, \(-\)7	&		&	C97	\\
351.445+0.660	&	17 20 54.61	&	 \(-\)35 45 08.6	&	120	&	\(-\)9.2	&	 \(-\)14, +1	&		&	C97	\\
351.581\(-\)0.353	&	17 25 25.12	&	 \(-\)36 12 46.1	&	44	&	\(-\)94.4	&	 \(-\)97, \(-\)92	&	c98	&	C97	\\
351.581\(-\)0.353n	&	17 25 25.18	&	 \(-\)36 12 44.5	&	2.3	&	\(-\)91.1	&	 \(-\)100, \(-\)88	&	 	&	C97	\\
351.775\(-\)0.536	&	17 26 42.57	&	 \(-\)36 09 17.6	&	230	&	1.3	&	 \(-\)9, +3	&	c98	&	CVF95, C97	\\
352.083+0.167	&	17 24 41.22	&	 \(-\)35 30 18.6	&	1.7	&	 \(-\)66.0	&	 \(-\)68, \(-\)64	&		&	99may	\\
352.111+0.176	&	17 24 43.56	&	 \(-\)35 28 38.4	&	8	&	 \(-\)54.8	&	 \(-\)61, \(-\)53	&		&	99may	\\
352.133\(-\)0.944	&	17 29 22.23	&	 \(-\)36 05 00.2	&	17.2	&	 \(-\)16.0	&	 \(-\)19, \(-\)6	&		&	99may	\\
352.517\(-\)0.155	&	17 27 11.34	&	 \(-\)35 19 32.3	&	6.3	&	 \(-\)51.2	&	 \(-\)52, \(-\)49	&	c98	&	97may	\\
352.525\(-\)0.158	&	17 27 13.42	&	 \(-\)35 19 15.5	&	0.75	&	\(-\)53.0	&	 \(-\)62, \(-\)52	&		&	97may	\\
352.630\(-\)1.067	&	17 31 13.91	&	 \(-\)35 44 08.7	&	160	&	 \(-\)2.8	&	 \(-\)7, \(-\)2	&	c98	&	96dec	\\
352.624\(-\)1.077	&	17 31 15.31	&	 \(-\)35 44 47.7	&	35	&	 +5.8	&	 \(-\)2, +7	&		&	96dec	\\
353.273+0.641	&	17 26 01.59	&	 \(-\)34 15 14.6	&	24	&	\(-\)5.2	&	 \(-\)6, \(-\)2	&		&	96dec	\\
353.410\(-\)0.360	&	17 30 26.18	&	 \(-\)34 41 45.6	&	86.8	&	\(-\)19.9	&	 \(-\)23, \(-\)19	&	c98	&	CVF95, C97	\\
353.464+0.562	&	17 26 51.53	&	 \(-\)34 08 25.7	&	18	&	\(-\)50.7	&	 \(-\)53, \(-\)49	&	c98	&	97may	\\
354.615+0.472	&	17 30 17.09	&	 \(-\)33 13 55.1	&	151	&	\(-\)24.6	&	 \(-\)27, \(-\)13	&	c98	&	CVF95, C97	\\
354.724+0.300	&	17 31 15.55	&	 \(-\)33 14 05.7	&	16	&	 +93.8	&	 +91, +95	&	c98	&	C97	\\
355.344+0.147	&	17 33 29.07	&	 \(-\)32 47 58.6	&	7.6	&	 +20	&	 +19, +21	&	c98	&	C97	\\
355.343+0.148	&	17 33 28.79	&	 \(-\)32 47 59.7	&	0.8	&	 +5.7	&	 +4, +7	&		&	C97	\\
355.346+0.149	&	17 33 28.92	&	 \(-\)32 47 49.0	&	9.2	&	        +10.0	&	 +9, +11	&		&	C97	\\
356.662\(-\)0.263	&	17 38 29.16	&	 \(-\)31 54 38.8	&	5	&	 \(-\)53.9	&	 \(-\)57, \(-\)44	&	c98	&	99feb	\\
357.967\(-\)0.163	&	17 41 20.26	&	 \(-\)30 45 06.9	&	35	&	 \(-\)3.2	&	 \(-\)6, 0	&	c98	&	97may	\\
357.965\(-\)0.164	&	17 41 20.14	&	 \(-\)30 45 14.4	&	1.5	&	 \(-\)8.8	&	 \(-\)9, +3	&		&	97may	\\
358.263\(-\)2.061	&	17 49 37.63	&	 \(-\)31 29 18.0	&	10	&	 +4.9	&	 +1, +6	&		&	99oct	\\
358.371\(-\)0.468	&	17 43 31.95	&	 \(-\)30 34 10.7	&	29	&	 +1.0	&	 \(-\)1, +13	&		&	97may	\\
358.386\(-\)0.483	&	17 43 37.83	&	 \(-\)30 33 51.1	&	2.5	&	 \(-\)6.0	&	 \(-\)7, \(-\)5	&	c98	&	97may	\\
359.138+0.031	&	17 43 25.67	&	 \(-\)29 39 17.3	&	15.6	&	\(-\)3.9	&	 \(-\)7, +1	&	c98	&	C96b, C97	\\
359.436\(-\)0.104	&	17 44 40.60	&	 \(-\)29 28 16.0	&	26.8	&	\(-\)52.0	&	 \(-\)53, \(-\)45	&	c98	&	C96b 	\\
359.436\(-\)0.102	&	17 44 40.21	&	 \(-\)29 28 12.5	&	4.4	&	\(-\)53.6	&	 \(-\)58, \(-\)54	&		&	C96b	\\
359.615\(-\)0.243	&	17 45 39.09	&	 \(-\)29 23 30.0	&	89	&	 +22.5	&	 +14, +27	&	c98	&	C96b	\\
359.970\(-\)0.457	&	17 47 20.17	&	 \(-\)29 11 59.4	&	1.3	&	 +23.0	&	 +20, +24	&	c98	&	C96b	\\
0.212\(-\)0.001	&	17 46 07.63	&	 \(-\)28 45 20.9	&	3.5	&	 +49.2	&	 +41, +50	&		&	C96b	\\
0.315\(-\)0.201	&	17 47 09.13	&	 \(-\)28 46 15.7	&	41.2	&	 +18	&	 +14, +27	&		&	C96b	\\
0.316\(-\)0.201	&	17 47 09.33	&	 \(-\)28 46 16.0	&	1.3	&	 +21	&	 +20, +22	&		&	C96b	\\
0.376+0.040	&	17 46 21.41	&	 \(-\)28 35 40.0	&	0.7	&	 +37.1	&	 +35, +40	&	c98	&	99feb	\\
0.475\(-\)0.010	&	17 46 47.05	&	 \(-\)28 32 07.1	&	2.9	&	 +28.7	&	 +23, +30	&		&	C96b	\\
0.496+0.188	&	17 46 03.96	&	 \(-\)28 24 52.8	&	10	&	 +0.8	&	 \(-\)12, +2	&	c98	&	C96b	\\
0.546\(-\)0.852	&	17 50 14.35	&	 \(-\)28 54 31.1	&	50	&	 +13.8	&	 +8, +20	&	c98	&	97may	\\
0.645\(-\)0.042	&	17 47 18.67	&	 \(-\)28 24 24.8	&	65	&	 +49.1	&	 +46, +53	&		&	C96b	\\
0.647\(-\)0.055	&	17 47 22.07	&	 \(-\)28 24 42.3	&	3.4	&	 +51.0	&	 +49, +52	&		&	C96b	\\
0.651\(-\)0.049	&	17 47 21.13	&	 \(-\)28 24 18.1	&	31.7	&	 +48.0	&	 +46, +49	&		&	C96b	\\
0.657\(-\)0.041	&	17 47 20.08	&	 \(-\)28 23 47.1	&	3	&	 +52.0	&	 +48, +56	&	text	&	C96b	\\
0.665\(-\)0.036	&	17 47 20.04	&	 \(-\)28 23 12.8	&	2.1	&	 +60.4	&	 +58, +62	&	text	&	C96b	\\
0.666\(-\)0.029	&	17 47 18.64	&	 \(-\)28 22 54.6	&	33.7	&	 +72.2	&	 +68, +73	&	text	&	C96b, C97	\\
0.672\(-\)0.031	&	17 47 20.03	&	 \(-\)28 22 41.7	&	4.5	&	 +58.2	&	 +55, +59	&	text	&	C96b	\\
0.677\(-\)0.025	&	17 47 19.29	&	 \(-\)28 22 14.6	&	4.4	&	 +73.4	&	 +70, +77	&		&	C96b	\\
0.695\(-\)0.038	&	17 47 24.76	&	 \(-\)28 21 43.2	&	26	&	 +68.5	&	 +64, +75	&		&	C96b	\\
0.836+0.184	&	17 46 52.86	&	 \(-\)28 07 34.8	&	8.1	&	 +3.5	&	 +2, +5	&		&	C96b	\\
2.143+0.009	&	17 50 36.14	&	 \(-\)27 05 46.5	&	6	&	 +62.7	&	 +55, +65	&	c98	&	96dec	\\
\hline
\end{tabular}
\medskip\\
\end{center}
\end{table*}

\begin{table*}[h]
\addtocounter{table}{-1}
\begin{center}
\caption{\textit{- continued p7 of 8}}\label{tableexample}
\begin{tabular}{lllrrccl}
\hline Galactic name & RA(2000) & Dec(2000) & I(pk) & v(pk) &
v(range) & OH? & refs, epoch $^a$ \\
\ (~~l,~~~~~~~b) & (h~~m~~~s) & ~~$^\circ$~~~ '~~~~" & (Jy) &
(\kms)
& (\kms) & & \\
\hline
2.536+0.198	&	17 50 46.47	&	 \(-\)26 39 45.3	&	40	&	 +3.2	&	 +2, +20	&		&	00nov	\\
3.910+0.001	&	17 54 38.75	&	 \(-\)25 34 44.8	&	2.4	&	 +17.8	&	 +17, +24	&	c98	&	97may	\\
5.900\(-\)0.430	&	18 00 40.86	&	 \(-\)24 04 20.8	&	5.3	&	 +10.0	&	 +4, +11	&		&	99oct	\\
6.539\(-\)0.108	&	18 00 50.86	&	 \(-\)23 21 29.8	&	0.9	&	 +13.4	&	 +13, +14	&		&	00nov	\\
6.610\(-\)0.082	&	18 00 54.03	&	 \(-\)23 17 02.1	&	10.8	&	 +0.7	&	 0, +1	&		&	00nov	\\
6.795\(-\)0.257	&	18 01 57.75	&	 \(-\)23 12 34.9	&	37	&	 +26.6	&	 +12, +31	&	c98	&	96dec	\\
8.139+0.226	&	18 03 00.75	&	 \(-\)21 48 09.9	&	3.5	&	  +20.0	&	 +19, +21	&		&	00nov	\\
8.669\(-\)0.356	&	18 06 18.99	&	 \(-\)21 37 32.2	&	8.5	&	 +39.3	&	 +39, +40	&	c98	&	96dec	\\
8.683\(-\)0.368	&	18 06 23.49	&	 \(-\)21 37 10.2	&	70	&	 +42.9	&	 +40, +46	&	c98	&	96dec	\\
9.621+0.196	&	18 06 14.67	&	 \(-\)20 31 32.4	&	5000	&	 +1.3	&	 \(-\)4, +9	&	c98	&	96dec	\\
9.619+0.193	&	18 06 14.92	&	 \(-\)20 31 44.3	&	72	&	 +5.5	&	 +5, +7	&	c98	&	96dec	\\
9.986\(-\)0.028	&	18 07 50.12	&	 \(-\)20 18 56.5	&	28	&	 +47.1	&	 +40, +52	&		&	00nov	\\
10.287\(-\)0.125	&	18 08 49.36	&	 \(-\)20 05 59.0	&	27	&	 +5.0	&	 +4, +6	&		&	00nov	\\
10.299\(-\)0.146	&	18 08 55.54	&	 \(-\)20 05 57.5	&	3.5	&	 +20.0	&	 +19, +21	&		&	00nov	\\
10.320\(-\)0.259	&	18 09 23.30	&	 \(-\)20 08 06.9	&	7.5	&	 +38.8	&	 +35, +40	&		&	00nov	\\
10.323\(-\)0.160	&	18 09 01.46	&	 \(-\)20 05 07.8	&	126	&	 +10.0	&	 +4, +14	&		&	00nov	\\
10.342\(-\)0.142	&	18 08 59.99	&	 \(-\)20 03 35.4	&	12	&	 +14.8	&	 +6, +18	&		&	00nov	\\
10.444\(-\)0.018	&	18 08 44.88	&	 \(-\)19 54 38.3	&	14.8	&	 +73.2	&	  +68, +79	&	c98	&	CVF95	\\
10.473+0.027	&	18 08 38.20	&	 \(-\)19 51 50.1	&	120	&	 +75.0	&	 +58, +77	&	c98	&	CVF95	\\
10.480+0.033	&	18 08 37.88	&	 \(-\)19 51 16.0	&	9.7	&	 +65.0	&	 +58, +66	&	c98	&	CVF95	\\
10.627\(-\)0.384	&	18 10 29.22	&	 \(-\)19 55 41.1	&	3.1	&	 +4.6	&	 \(-\)6, +7	&		&	98nov	\\
10.629\(-\)0.333	&	18 10 17.98	&	 \(-\)19 54 04.8	&	4.2	&	 \(-\)7.5	&	 \(-\)13, +1	&		&	98nov	\\
10.958+0.022	&	18 09 39.32	&	 \(-\)19 26 28.0	&	16.4	&	 +24.4	&	 +23, +26	&		&	00nov	\\
11.034+0.062	&	18 09 39.84	&	 \(-\)19 21 20.3	&	0.5	&	 +20.6	&	 +15, +21	&	c98	&	98nov	\\
11.497\(-\)1.485	&	18 16 22.13	&	 \(-\)19 41 27.1	&	167	&	 +6.7	&	 +4, +17	&		&	00nov	\\
11.904\(-\)0.141	&	18 12 11.44	&	 \(-\)18 41 28.6	&	56	&	 +42.8	&	 +40, +45	&	c98	&	C97	\\
11.903\(-\)0.102	&	18 12 02.70	&	 \(-\)18 40 24.7	&	1.8	&	 +36.0	&	 +32, +37	&		&	C97	\\
11.936\(-\)0.150	&	18 12 17.29	&	 \(-\)18 40 02.6	&	1.9	&	 +48.5	&	 +47, +50	&		&	C97	\\
11.936\(-\)0.616	&	18 14 00.89	&	 \(-\)18 53 26.6	&	41	&	 +32.2	&	 +30, +44	&		&	00nov	\\
12.025\(-\)0.031	&	18 12 01.86	&	 \(-\)18 31 55.7	&	85	&	 +107.7	&	 +105, +112	&	text	&	00nov	\\
12.209\(-\)0.102	&	18 12 39.92	&	 \(-\)18 24 17.9	&	9.2	&	 +19.8	&	 +16, +22	&	text	&	97may	\\
12.202\(-\)0.120	&	18 12 42.93	&	 \(-\)18 25 11.8	&	1.7	&	 +26.4	&	 +26, +27	&		&	97may	\\
12.203\(-\)0.107	&	18 12 40.24	&	 \(-\)18 24 47.5	&	2.4	&	 +20.5	&	 +20, +32	&		&	97may	\\
12.181\(-\)0.123	&	18 12 41.00	&	 \(-\)18 26 21.9	&	1.35	&	 +29.7	&	 +29, +31	&		&	97may	\\
12.199\(-\)0.034	&	18 12 23.44	&	 \(-\)18 22 50.9	&	12.5	&	 +49.3	&	 +48, +53	&		&	97may	\\
12.265\(-\)0.051	&	18 12 35.40	&	 \(-\)18 19 52.3	&	6	&	 +68.0	&	 +58, +70	&		&	97may	\\
12.625\(-\)0.017	&	18 13 11.30	&	 \(-\)17 59 57.6	&	11.2	&	 +23.8	&	 +21, +28	&		&	00nov	\\
12.681\(-\)0.182	&	18 13 54.75	&	 \(-\)18 01 46.6	&	450	&	 +57.6	&	 +50, +61	&	c98	&	98nov	\\
12.889+0.489	&	18 11 51.40	&	 \(-\)17 31 29.6	&	27	&	 +39.3	&	 +28, +42	&	c98	&	96dec	\\
12.909\(-\)0.260	&	18 14 39.53	&	 \(-\)17 52 00.0	&	300	&	 +39.9	&	 +35, +47	&	c98	&	98nov	\\
13.657\(-\)0.599	&	18 17 24.27	&	 \(-\)17 22 12.5	&	32	&	 +51.2	&	 +47, +53	&	text	&	05mar	\\
14.101+0.087	&	18 15 45.81	&	 \(-\)16 39 09.4	&	146	&	 +15.2	&	 +4, +17	&		&	00nov	\\
15.034\(-\)0.677	&	18 20 24.78	&	 \(-\)16 11 34.6	&	28	&	 +21.2	&	 +20, +24	&	c98	&	C97	\\
17.638+0.157	&	18 22 26.30	&	 \(-\)13 30 12.1	&	9.5	&	 +20.7	&	 +20, +22	&	c04	&	2003mar	\\
19.472+0.170	&	18 25 54.70	&	 \(-\)11 52 34.6	&	8.5	&	 +21.7	&	 +17, +23	&		&	99oct	\\
19.472+0.170sw	&	18 25 54.49	&	 \(-\)11 52 36.5	&	1.7	&	 +13.7	&	 +13, +14	&		&	99oct	\\
19.486+0.151	&	18 26 00.39	&	 \(-\)11 52 22.6	&	12.5	&	 +20.9	&	 +19, +25	&		&	99oct	\\
19.496+0.115	&	18 26 09.16	&	 \(-\)11 52 51.7	&	5.3	&	 +121.0	&	 +120, +122	&		&	99oct	\\
20.237+0.065	&	18 27 44.56	&	 \(-\)11 14 54.2	&	77	&	 +71.8	&	 +68, +78	&	c03	&	99oct	\\
20.239+0.065	&	18 27 44.95	&	 \(-\)11 14 48.9	&	5.5	&	 +70.6	&	 +60, +71	&		&	99oct	\\
\hline
\end{tabular}
\medskip\\
\end{center}
\end{table*}

\begin{table*}[h]
\addtocounter{table}{-1}
\begin{center}
\caption{\textit{- continued p8 of 8}}\label{tableexample}
\begin{tabular}{lllrrccl}
\hline Galactic name & RA(2000) & Dec(2000) & I(pk) & v(pk) &
v(range) & OH? & refs, epoch $^a$ \\
\ (~~l,~~~~~~~b) & (h~~m~~~s) & ~~$^\circ$~~~ '~~~~" & (Jy) &
(\kms)
& (\kms) & & \\
\hline
22.335\(-\)0.155	&	18 32 29.40	&	 \(-\)09 29 30.1	&	43	&	 +35.7	&	 +35, +37	&		&	99oct	\\
22.435\(-\)0.169	&	18 32 43.82	&	 \(-\)09 24 33.2	&	16	&	 +29.3	&	 +22, +40	&		&	99oct	\\
23.010\(-\)0.411	&	18 34 40.27	&	 \(-\)09 00 38.3	&	300	&	 +75.0	&	 +72, +83	&		&	99feb	\\
23.437\(-\)0.184	&	18 34 39.25	&	 \(-\)08 31 38.5	&	45	&	 +103.0	&	 +101, +108	&		&	99feb	\\
23.440\(-\)0.182	&	18 34 39.18	&	 \(-\)08 31 24.3	&	25	&	 +96.6	&	 +94, +100	&		&	99feb	\\
24.329+0.144	&	18 35 08.09	&	 \(-\)07 35 03.6	&	7	&	 +110.5	&	 +109, +120	&		&	05oct	\\
24.493\(-\)0.039	&	18 36 05.83	&	 \(-\)07 31 20.6	&	12	&	 +115.0	&	 +109, +116	&		&	05oct	\\
28.146\(-\)0.005	&	18 42 42.59	&	 \(-\)04 15 36.5	&	61	&	 +101.2	&	 +99, +105	&	text	&	99oct	\\
28.201\(-\)0.049	&	18 42 58.08	&	 \(-\)04 13 56.2	&	3.5	&	 +98.9	&	 +94, +100	&	text	&	99oct	\\
40.425+0.700	&	19 02 39.62	&	 +06 59 10.5	&	15	&	 +15.7	&	 +5, +16	&		&	99oct	\\
40.623\(-\)0.138	&	19 06 01.63	&	 +06 46 36.5	&	12.5	&	 +31.1	&	 +29, +37	&		&	99oct	\\
43.149+0.013	&	19 10 11.06	&	 +09 05 20.0	&	18	&	 +13.5	&	 +13, +15	&		&	99oct	\\
43.165+0.013	&	19 10 12.89	&	 +09 06 11.9	&	26	&	 +9.3	&	 +7, +21	&		&	99oct	\\
43.171+0.004	&	19 10 15.36	&	 +09 06 15.2	&	10	&	 +20.2	&	 +18. +22	&		&	99oct	\\
43.167\(-\)0.004	&	19 10 16.27	&	 +09 05 50.6	&	2.4	&	\(-\)1.1	&	 \(-\)2, 0	&		&	99oct	\\
43.796\(-\)0.127	&	19 11 53.97	&	 +09 35 53.5	&	50	&	 +39.5	&	 +39, +43	&		&	99oct	\\
45.467+0.053	&	19 14 24.15	&	 +11 09 43.0	&	4.1	&	 +56.4	&	 +55, +59	&		&	99oct	\\
45.473+0.134	&	19 14 07.36	&	 +11 12 15.7	&	6.1	&	 +65.5	&	 +65, +67	&		&	99oct	\\
45.493+0.126	&	19 14 11.35	&	 +11 13 06.2	&	10	&	 +57.1	&	 +56, +58	&		&	99oct	\\
45.445+0.069	&	19 14 18.31	&	 +11 08 59.4	&	0.84	&	 +50.0	&	 +49, +51	&		&	99oct	\\
49.470\(-\)0.371	&	19 23 37.90	&	 +14 29 59.4	&	9.2	&	 +64.0	&	 +63, +76	&		&	98nov, 99oct	\\
49.471\(-\)0.369	&	19 23 37.60	&	 +14 30 05.4	&	4.1	&	 +73.2	&	 +72, +75	&		&	99oct	\\
49.482\(-\)0.402	&	19 23 46.19	&	 +14 29 47.0	&	7	&	 +50.0	&	 +47, +52	&		&	99oct	\\
49.489\(-\)0.369	&	19 23 39.83	&	 +14 31 05.0	&	26	&	 +56.1	&	 +55, +61	&		&	98nov, 99oct	\\
49.490\(-\)0.388	&	19 23 43.95	&	 +14 30 34.2	&	650	&	 +59.2	&	 +51, +61	&		&	98nov, 99oct	\\
\hline				
\end{tabular}
\medskip\\
$^a$refs abbreviated in the format: CVF95 is Caswell, Vaile 
\& Forster 1995;  C97 is Caswell 1997 etc.;  epochs abbreviated as year 
and month.  \\
\end{center}
\end{table*}

For the current analysis. we do not list each detected maser spot, 
but we do list, as separate, the site of each compact group of maser spots 
that is separated more than a few arcsec from any other group in the same 
cluster.  In a few rare cases there is a cluster of three or more groups; 
somewhat more commonly there is a pair, and the majority of sites are 
single.

The emphasis of the current work is to provide a catalogue of precise 
positions.  In order to increase its usefulness, Table 1 lists not only 
our new previously unpublished positions, but also: the 19 sources earlier 
listed only with 
1950 coordinates by Caswell Vaile and Forster (1995b), results from 
Caswell 1996a and 1996b (80 sources), and results from Caswell 
1997 (42 sources).  In addition, some of our new positions, although not 
published in tabular form, have been referred to in source notes 
concerning related OH 
masers at 1665 MHz (Caswell 1998) and at 6035 MHz (Caswell 2003). For some 
sources, we made new measurements confirming or replacing the previous 
published values.  The present positions are recommended for future 
studies and, unless otherwise noted, have rms uncertainties of 0.4 arcsec.  
These errors arise chiefly from residual atmospheric phase instabilities 
between calibration measurements as discussed by Caswell (1997).  

The peak intensity and its velocity at our observing epoch is shown in 
columns 4 and 5 of Table 1.  However, variability occurs in many sources, 
primarily on timescales of months to years (see e.g. Caswell, Vaile \& 
Ellingsen 1995c).  Consequently, the peak intensity at our observing epoch 
is often different from that of published spectra (and sometimes a 
different velocity peak is stronger).  Earlier published spectra (e.g. 
Caswell et al. 1995a) sometimes have a better signal-to-noise ratio than 
our measurements, and show features over a larger velocity range.  In 
column 6 we quote the larger ranges in such cases if there 
is no evidence that any emission comes from an offset position.  The 
resulting velocity ranges are approximate, and are generally 
underestimated for weak sources with low signal-to-noise ratio, 
but occasionally overestimated if two sources are blended.  A better 
assessment of velocity ranges will be possible from spectra being 
obtained in a new methanol multibeam survey, which is now well under way 
(Green et al 2009a).

The Table assigns a name to each maser based on its Galactic coordinates 
(to the nearest millidegree). 
Where separations of approximately 2 arcsec occur, it is not clear whether 
the features represent distinctly separate maser sites, or an unusually 
extended one.  The difficult cases are discussed for individual sources in 
the next section.

The methanol masers listed here include the results of searches towards a 
comprehensive list of southern OH masers (Caswell 1998) in the Galactic 
longitude range  232$^\circ$ through 360$^\circ$ to 16$^\circ$. 
Some of the methanol masers detected towards OH targets are coincident 
with the OH and others are not.  Coincidences are  
identified in the column `OH?' by citing a reference to the OH 
data;  most references are to Caswell (1998), but there are others 
positioned more recently.  Some associations require more extensive 
comment, and 'text' in the `OH?' column refers to notes in section 4.  
The resulting detection statistics of methanol masers towards 1665-MHz 
masers are discussed later in the paper.  

The final column of the Table identifies the epoch of our methanol 
position measurement or, where the measurement has been discussed in an 
earlier publication, a reference is given.  Published positions have 
sometimes been improved by additional data at a later epoch.  

There are, of course, multiple individual publications on many previously 
reported sources but we have not attempted to cite these since they are of 
varying quality and are 
mostly included in a comprehensive compilation of older methanol maser 
data (Pestalozzi Minier and Booth 2005).  Many of the positions 
listed here supersede the earlier approximate positions, and others 
confirm independently obtained positions of high accuracy.  
Amongst the positions in the Pestalozzi compilation with accuracy 
comparable to ours, those obtained by Walsh et al. (1998) were derived 
from ATCA observations in 1994 and 1995, and used a strategy similar to 
the present one (but with lower spectral resolution). For most of those 
sources, the strongest features, tabulated by Pestalozzi et al. 
(2005) from the full Walsh et al. (1998) dataset, are in agreement with 
our values to within 0.4 arcsec and provide a useful corroboration of both 
datasets.

Some northern sources have recently been measured with the Arecibo 
telescope, with an rms position uncertainty of 7 arcsec (Pandian, 
Goldsmith \& Deshpande 2007).  Comparison with our data confirms this 
precision, but also reveals a bias in the Arecibo Right Ascensions, 
suggesting that the values should be reduced by an average of 0.6s       
(9 arcsec).

\section{Discussion}

\subsection{Notes on some individual sources}

We first draw attention to corrections needed for earlier published data.  
A source listed by Caswell 
(1996a) as 335.603-0.078 is now believed to be spurious and is 
accounted for as a weak distant sidelobe of another maser.  
The site listed here as 0.475-0.010 is the corrected value for a source 
listed by Caswell (1996b) as 0.393-0.034.  

Problems of this type can occur when sparse antenna arrays are used to 
observe weak sources for only short periods, but such errors are rare and 
it is expected that no similar examples remain in the current catalogue.

The remaining notes draw attention to some anomalies regarding the 
information on a few of the sites, and draw attention to sites with 
neighbours less than 20 arcsec away, which mostly represent individual 
stars within a cluster.  A few of the sites are as close as a few 
arcsec and the alternative interpretations of two close 
separate sites, or a single site of larger extent, are discussed.

\subparagraph{291.579-0.431 and 291.582-0.435.} As noted previously 
(Caswell 2004a) these two sites, with separation nearly 20 arcsec, are 
sufficiently close to lie within the same star cluster, but are 
quite distinct, with one site accompanied by water, and the other 
accompanied by both water and OH.  

\subparagraph{311.947+0.472.} This has a possibly associated OH maser 
but the OH position still has an uncertainty larger than 10 arcsec 
(Caswell 1998).  

\subparagraph{312.598+0.045 and 312.597+0.045.} The first of these sites
is stronger and coincides with an OH maser.  The second methanol site is
offset 6 arcsec and appears to be distinct, and has no detected OH
counterpart.

\subparagraph{319.836-0.197.}  The apparent OH counterpart is offset 3 
arcsec, but is weak, and its position uncertainty may account for the 
discrepancy;  we provisionally regard the two species of maser as 
coincident.  

\subparagraph{321.030-0.485 and 321.033-0.483.} The separation is more 
than 10 arcsec and only the first, weaker, site has a detected OH 
counterpart. 

\subparagraph{327.392+0.199 and 327.395+0.197.}  The separation of these 
sources is 14 arcsec and neither has a detected OH counterpart.  

\subparagraph{328.808+0.633 and 328.809+0.633}  These were treated by 
Caswell (1997) as possibly separate sites, partly on the basis of an 
overlay with continuum emission, and despite their small separation of 
only 2.4 arcsec.  The spectral features of the second source lie within 
the velocity range of those from the first source, but we retain them as 
distinct sources pending more evidence.  

\subparagraph{329.339+0.148.}  The discovery of the 1665-MHz maser at 
this site was reported by Caswell (2001) and it turns out to be an 
especially interesting distant site with OH maser emission also at 1720 
MHz (Caswell 2004b) and 13441 MHz (Caswell 2004c).  

\subparagraph{329.405-0.459 and 329.407-0.459.} The separation of the 
sites is 5.7 arcsec and only the first site coincides with an OH maser.  

\subparagraph{330.953-0.182.}  As noted by Caswell (2001), the methanol 
coincides with an OH 6035-MHz maser site, but the major OH 1665 and 
1667-MHz masers are offset 3 arcsec.  

\subparagraph{331.542-0.066 and 331.543-0.066.} The separation is only 3 
arcsec, but at a likely distance of 6 kpc this correponds to 90 mpc, 
and there is further evidence that they represent two distinct sites 
since there is no obvious velocity overlap, and each has an OH 
counterpart (Caswell 1997, 1998).  

\subparagraph{333.126-0.440 and 333.128-0.440.} The separation is 7 
arcsec and OH emission has not been detected at either site.  

\subparagraph{333.135-0.431s.} The suffix denoting south was added by 
Caswell (1997) to distinguish this site from another site, offset nearly 
3 arcsec, which has OH without methanol.

\subparagraph{333.128-0.560 and 333.130-0.560.}  The sites are distinct 
with separation more than 6 arcsec and neither has a detected OH 
counterpart.

\subparagraph{335.585-0.289 and 335.585-0.290.}  These appear to 
represent two distinct sites, but with a separation of barely 3 arcsec, 
this is uncertain.  The first site has coincident OH.  The second is a 
single strong feature at an offset velocity and without OH emission.  

\subparagraph{337.703-0.053 and 337.705-0.053.}  Only the second, 
stronger, maser of this pair has a detected OH counterpart. 

\subparagraph{338.075+0.012 and 338.075+0.009.} Only the first, stronger, 
maser of this pair has a detected OH counterpart. 

\subparagraph{339.681-1.208 and 339.682-1.207.}  As noted by Caswell 
(1998) there is an OH counterpart to the second site straddling the 
methanol position.  The first methanol site which lies more than 3 arcsec 
south is stronger, has a wide velocity range encompassing the small range 
of the second source and although it seems to be spatially distinct, this 
is not certain.  

\subparagraph{344.419+0.044 and 344.421+0.045.}  These are clearly 
distinct and only the first, weaker, methanol site has an OH counterpart. 

\subparagraph{345.003-0.223 and 345.003-0.224.}  The second of these 
agrees well in position with an OH counterpart at both the 1665- and 
6035-MHz transitions.  The velocity ranges of the two sources do 
not noticeably overlap and an overlay on continuum emission argues 
strongly that the two sites are quite distinct despite their small 
separation of only 3 arcsec (Caswell 1997).

\subparagraph{345.010+1.792 and 345.012+1.797.}  These are clearly 
distinct and only the first has a detected OH counterpart.

\subparagraph{348.550-0.979 and 348.550-0.979n.}  The first of these has 
a continuum and OH counterpart at both the 1665- and 6035-MHz transitions.  
There is evidence (Caswell 1997) that the second source (offset 2.2 
arcsec to the north), detected only on the methanol transition, is quite 
likely a site with its own source of excitation.  The sites have 
overlapping velocity ranges.  

\subparagraph{349.092+0.105 and 349.092+0.106.} The second source 
coincides with an OH maser.  The first source, with no OH, has a clearly 
distinct velocity range and seems to be a separate site despite 
the smallness of its offset, slightly more than 2 arcsec, from the second.  

\subparagraph{350.105+0.083 and 350.104+0.084.}  Separated nearly 4 
arcsec, neither site has an OH counterpart.  The small velocity range of 
the second source lies wholly within the range of the first.  

\subparagraph{351.417+0.645 and 351.417+0.646.}  The separation of these 
sites is more than 3 arcsec.  The first one coincides with the 
well-known \HII\ region NGC6334F and its OH maser emission.  The second 
lies clearly offset and appears to be a distinct separate site (Caswell 
1997).  

\subparagraph{351.581-0.353.}  Caswell (1997) 
queried whether there was an additional distinct source to the north 
(351.581-0.353n).  The separation of slighly less than 1.8 arcsec leaves 
this unclear, and the extra maser spots have velocities within the range 
of the main site.    
We list the positions of both sites, but note that the 1665-MHz 
OH counterpart and a compact \HII\ region lie between the positions, which 
hints that we are more likely dealing with an extended single site.  
Recent recognition that the site is most likely in the near portion of the 
expanding 3-kpc arm (Green et al. 2009b) would imply a distance of 5.3 
kpc, and thus a linear extent of 45 mpc which, for a single site, is large 
but not exceptional.  

\subparagraph{353.273+0.641.}  This strong methanol maser was first 
detected 1993 but was reported only recently (Caswell \& Phillips 2008) 
when an association with a remarkable water maser was confirmed.  No OH 
maser has been detected at the site.  

\subparagraph{355.344+0.147, 355.343+0.148.}  A third 
maser, 355.346+0.149, is quite distinct spatially, with an 
offset of 10 arcsec from the other sites.  The first two sites have 
no overlap in velocity but are separated spatially only 3.7 arcsec, and 
the likelihood that they are distinct sites depends on an estimate of 
their distance.  Crovisier, Fillit and Kazes (1973) argue, on the basis of 
intervening absorption features near +100 \kms, that the radio continuum 
emission and the OH 1665-MHz maser emission are at a distance beyond the 
Galactic Centre.  This evidence subsequently passed unnoticed in the 
literature, attracting no relevant citations, and in particular 
was overlooked by Caswell (1997) and Forster \& Caswell (1989, 1999, 2000) 
who assigned the complex to a distance of only 2 kpc.  If we reject the 
nearby location, the alternative distances then include:  outside the 
solar circle and thus beyond 17 kpc;  near the Galactic Centre at 8.5 kpc 
which might account for unusual velocities;  or (perhaps most likely) a 
location in the far-side counterpart to the 3-kpc expanding arm at 11.5 
kpc (Dame \& Thaddeus 2008;  Green et al. 2009b).   At any distance beyond 
8.5 kpc, the separation of 3.7 arcsec then corresponds to more than 150 
mpc, indicative of clearly distinct sites for all three methanol masers.  

\subparagraph{357.967-0.163 and 357.965-0.164.}  Two clearly distinct 
sites, separated 7.6 arcsec, of which only the first has an OH 
counterpart.  

\subparagraph{359.436-0.104 and 359.436-0.102.}  Clearly distinct 
and separated 6 arcsec, the first has a well-known OH counterpart (Caswell 
1998) and the second also now has a more recently reported OH maser 
counterpart (Argon et al. 2000).  

\subparagraph{0.315-0.201 and 0.316-0.201}  These are  separated by 
2.5 arcsec and listed as distinct sites by  Caswell (1996b), despite the  
weak features of the second lying wholly within the range of the first 
source.  At neither position is there any detected OH counterpart. Pending 
further evidence, we list both sites, but caution that they may in fact be 
components of a more than usually extended single site.  

\subparagraph{0.475-0.010.} This source was discovered as a new source in 
the Galactic centre survey by Caswell (1996b) but was incorrectly reported 
as 0.393-0.034.  Re-analysis of the data showed that the position error 
arose because the sparse uv-coverage caused a sidelobe to be of comparable 
amplitude to the main lobe and was incorrectly interpreted as the source 
position.  The source is closer to the target pointing than first 
estimated, and so its flux 
density correction for the offset is not as large, and the new estimate of 
peak flux density is therefore lower, 2.9 Jy.  

\subparagraph{0.645-0.042 to 0.695-0.038.}  These nine sites within 
the Sgr B2 complex were distinguished by both Houghton and Whiteoak (1995)  
and by Caswell (1996b); they are all clearly distinct sites. 
Existing OH 1665 and 1667-MHz observations towards Sgr B2 remain 
incomplete, and those by Argon et al. (2000) are some of the best 
currently available.  The detailed information in their datasets show 
counterparts at 1665 and 1667 MHz for 0.657-0.041 and 0.672-0.031, and an 
OH 1720 MHz counterpart for 0.665-0.036.  
One of the other methanol sites, 0.666-0.029, is accompanied by a 6035-MHz 
maser (Caswell 1997).    

Two weak additional methanol sites in the Sgr B2 complex were reported by 
Houghton and Whiteoak (1995) and are believed reliable but were too weak 
to confirm in the present observations.  They are omitted from the present 
listing which is intended to present only the results of our independent 
observations.  

\subparagraph{9.621+0.196 and 9.619+0.193.}  The first of these is the 
strongest known methanol maser, and the second is a clearly distinct site 
offset more than 10 arcsec.  Both have OH counterparts and uc\HII\ 
regions (Forster \& Caswell 2000).  

\subparagraph{11.034+0.062.} Note that the weak feature seen on the 
spectrum of Caswell et al. (1995a) at velocity 24.4 \kms\ is a sidelobe of 
10.958+0.022.  

\subparagraph{12.025-0.031.}  There is an OH 1665-MHz maser counterpart at 
18$^{h}$12$^{m}$01.88$^{s}$, -18$^{\circ}$31'55.6" (Caswell unpublished; 
this is a precise position for a source previously listed as 12.03-0.04 by 
Caswell 1998) and it is thus now confirmed to coincide with the methanol. 

\subparagraph{12.209-0.102.}  A 1665-MHz OH maser counterpart lies at this 
position (Argon et al. 2000; and Caswell unpublished).  

\subparagraph{12.889+0.489.}  Spectral features over the velocity range 28 
to 42 \kms\ mostly lie within 0.5 arcsec of the tabulated position, but a 
single strong feature at velocity +33.5 \kms\ is offset 2 arcsec 
northeast, at 
18$^{h}$11$^{m}$51.49$^{s}$, -17$^{\circ}$31'28.0". 
The OH counterpart is offset from both methanol features by slightly more 
than 1 arcsec and we treat this as a single site.    

\subparagraph{13.657-0.609.}  This source was first reported by MacLeod 
et al. (1998) but is not in the compilation of Pestalozzi et al. (2005). 
As also noted by MacLeod et al., there is associated OH emission at 1665 
and 1667 MHz which new observations (Caswell unpublished) show to be at  
18$^{h}$17$^{m}$24.27$^{s}$, -17$^{\circ}$22'13.4", effectively coincident 
with the methanol.  

\subparagraph{19.472+0.170 and 19.472+0.170sw.} The second source is 
offset 3.7 arcsec south-west from the first, and offset to smaller 
velocity.  No other maser is known at these positions so it is not clear 
whether the sites are distinct, or simply a larger than usual single site.  

\subparagraph{20.237+0.065 and 20.239+0.065.}  The two sites are separated 
more than 7 arcsec.  The first coincides with OH maser emission at 1665 
MHz, 6035 MHz and 1720 MHz (Caswell 2003, 2004b), whereas the second is 
solitary.  

\subparagraph{23.437-0.184 and 23.440-0.182.}  The clear separation of 
more than 10 arcsec establishes these as distinct sites with distinct 
velocity ranges.  

\subparagraph{28.146-0.005 and 28.201-0.049.} The proximity of these 
sources to declination zero causes the beamsize in declination to be 
large, and the declinations to have larger than usual uncertainties, 
estimated to be 1 arcsec.  The correspondence in each case with an OH 
maser (Argon et al. 2000) to better than 2 arcsec suggests that our 
errors are indeed no greater than 2 arcsec. 

\subparagraph{43.149+0.013 to 49.161+0.004.}  These four sites are part 
of the W49 complex and were noted as distinct in the single dish 
observations of Caswell et al. (1995a).  Pandian et al. (2007) detect all 
four, plus an additional weak one of less than 1 Jy peak.

\subparagraph{49.470-0.371 to 49.490-0.388.} These five sites are part of 
the complex W51.  Caswell et al (1995a) recognised that there were at 
least three sites here and Pandian et al. (2007) recognised four.  Our 
higher resolution now distinguishes five sites with clearly defined 
separate positions, although the velocity ranges of weak features are 
uncertain owing to sidelobe confusion.  

\subsection{Association with OH masers}

The precise methanol maser positions reported here allow an improved study 
of the association of methanol masers and 1665-MHz OH masers 
in regions of Massive Star Formation.  However, beyond the 16 degree 
Galactic longitude limit of the Caswell (1998) catalogue, the OH 
information is incomplete, although some individual sources can be studied 
using the OH positions available in Forster \& Caswell (1989, 1999) and in 
Argon et al. (2000).  

Therefore, for statistical purposes in the evaluation of the discovery 
statistics of methanol towards OH masers, we consider only the 
Galactic longitude range 232$^\circ$ through 360$^\circ$ to 16$^\circ$, 
covered by the OH catalogue of Caswell (1998), 
A preliminary analysis was performed by Caswell (1998), but accurate 
positions for some of the methanol masers were not known and some of the 
possible associations were therefore  uncertain.  Furthermore, there are 
several more recent OH results in this region as noted in section 4.1.  
We find that for the (updated) list of 207 Star Formation Region OH masers 
with precise positions known in this longitude range, 168 (81 per cent) 
possess a methanol counterpart.  
Of course, the interpretation of methanol and OH associations in 
terms of common conditions and evolutionary stages for methanol and OH 
co-existence requires a closer inspection of line ratios and investigation 
of co-propagation.

However, a practical consequence of this 81 per cent statistic is that 
when a new, deep, 
unbiassed survey for methanol masers has been completed (Green 2009a), 
and the positions used as targets for an OH search, we may expect the 
results to be a useful proxy for a deep unbiased survey for OH, and 
perhaps to recover at least 80 per cent of the full OH population.

\subsection{Unusually wide velocity spreads}

We have explored the velocity widths of the methanol masers and find 
that only nine of our sample have velocity widths exceeding 16 \kms. The 
largest is 24 \kms, shown by 340.785-0.096; its extreme red-shifted 
emission is very weak, at an intensity only 1 per cent of the strongest 
region of 
emission (Caswell et al. 1995a;  Caswell 1997).  The two sources with 
extent 17 \kms\ (344.227-0.569 and 
340.054-0.244) also have only weak emission at one of their extreme 
velocities, one blue, the other red.  
The six sources with velocity ranges of 18 or 19 \kms\ 
(339.681-1.208, 330.070+1.064, 10.473+0.027, 2.536+0.198, 6.795-0.257, 
and 22.435-0.155) have somewhat 
stronger emission near both extremes of velocity, the intensity ratio 
of blue to red ranging from 1.8 to 0.1.  

Since velocity extents greater than 16 \kms\ are rare (less than 3 per 
cent of the total), this lends validity to the practice of using methanol 
velocities (e.g. the mid-values of the range) as the systemic velocity, 
with the expectation that the uncertainty is rarely as large as 10 \kms, 
and most commonly less than 5 \kms.

The systemic velocity is dominated by Galactic rotation for the Galactic 
disk population of young massive stars, and is thus suitable for 
estimating kinematic distances.  Velocity ranges of individual sources 
often match those of OH counterparts quite well, and confirm the likely 
systemic velocities.

\section{Conclusion}

The precise positions reported here confirm that more than 80 percent of 
OH masers in regions of massive star formation have associated methanol 
masers.  

We have also explored the velocity widths of the methanol masers and find 
that values greater than 16 \kms\ are rare (less than three per cent 
of the total), in marked contrast to water masers where more than half the 
sources have velocity widths exceeding 20 \kms.  



\end{document}